\documentclass[sigconf]{acmart}

\copyrightyear{2024}
\acmYear{2024}
\setcopyright{acmlicensed}\acmConference[ASE '24]{39th IEEE/ACM
International Conference on Automated Software Engineering }{October
27-November 1, 2024}{Sacramento, CA, USA}
\acmBooktitle{39th IEEE/ACM International Conference on Automated Software
Engineering (ASE '24), October 27-November 1, 2024, Sacramento, CA, USA}
\acmDOI{10.1145/3691620.3694998}
\acmISBN{979-8-4007-1248-7/24/10}

\AtBeginDocument{%
  \providecommand\BibTeX{{%
    Bib\TeX}}}

\usepackage{subfig,graphicx}
\usepackage{enumitem}
\usepackage{bbding}
\usepackage{listings}
\usepackage{multirow}
\usepackage{makecell}
\usepackage{xcolor}
\usepackage{listings}
\usepackage{pifont}
\usepackage{xspace}
\makeatletter
\DeclareRobustCommand\bmvaOneDot{\futurelet\@let@token\bmv@onedotaux}
\def\bmv@onedotaux{\ifx\@let@token.,\else.\null\fi\xspace}
\DeclareRobustCommand\bmvaTwoDot{\futurelet\@let@token\bmv@twodotaux}
\def\bmv@twodotaux{\ifx\@let@token.,\else.,\null\fi\xspace}
\makeatother
\def\eg{\emph{e.g}\bmvaTwoDot} 
\def\ie{\emph{i.e}\bmvaTwoDot}

\def\wrt{w.r.t\bmvaOneDot} 
\def\etal{\emph{et al}\bmvaOneDot}

\newcommand{\InstrumentationToolname}{\textit{DLModelExplorer}\xspace}
\newcommand{\ObfuscationToolname}{\textit{DynaMO}\xspace}

\usepackage{tcolorbox}
\def\BibTeX{{\rm B\kern-.05em{\sc i\kern-.025em b}\kern-.08em
    T\kern-.1667em\lower.7ex\hbox{E}\kern-.125emX}}

\theoremstyle{plain}
\newtheorem{theorem}{Theorem}[section]
\newtheorem{lemma}[theorem]{Lemma}
\theoremstyle{definition}

\usepackage{amsmath}
\usepackage[capitalize,noabbrev]{cleveref}

\definecolor{codegreen}{RGB}{2,112,10}
\definecolor{codegray}{rgb}{0.5,0.5,0.5}
\definecolor{codepurple}{rgb}{0.58,0,0.82}
\definecolor{backcolour}{rgb}{0.95,0.95,0.92}

\usepackage{array} 
 
\lstdefinestyle{mystyle}{
  language=C++,
  backgroundcolor=\color{backcolour},   commentstyle=\color{codegreen},
  keywordstyle=\color{magenta},
  numberstyle=\tiny\color{codegray},
  stringstyle=\color{codepurple},
  basicstyle=\ttfamily\footnotesize,
  breakatwhitespace=false,         
  breaklines=true,                 
  captionpos=b,                    
  keepspaces=true,                 
  numbersep=5pt,                  
  showspaces=false,                
  showstringspaces=false,
  showtabs=false,                  
  tabsize=2
}

\begin{document}

\title{\ObfuscationToolname: Protecting Mobile DL Models through Coupling Obfuscated DL Operators}

\author{Mingyi Zhou} \authornote{This work was partially done when Mingyi Zhou was a PhD student at Monash University}
\orcid{0000-0003-3514-0372}
\affiliation{%
  \institution{Beihang University}
  \city{Beijing}
  \country{China}
}

\email{mingyi.zhou@monash.edu}

\author{Xiang Gao}
\orcid{0000-0001-9895-4600}
\affiliation{%
  \institution{Beihang University}
  \city{Beijing}
  \country{China}
}
\email{xiang_gao@buaa.edu.cn}

\author{Xiao Chen}
\orcid{0000-0002-4508-5971}
\affiliation{%
  \institution{University of Newcastle}
  \city{Callaghan}
  \country{Australia}
}
\email{xiao.chen@newcastle.edu.au}

\author{Chunyang Chen}
\orcid{0000-0003-2011-9618}
\affiliation{%
  \institution{TU Munich}
  \city{Heilbronn}
  \country{Germany}
}
\email{chun-yang.chen@tum.de}

\author{John Grundy}
\orcid{0000-0003-4928-7076}
\affiliation{%
  \institution{Monash University}
  \city{Clayton}
  \country{Australia}
}
\email{john.grundy@monash.edu}

\author{Li Li}\authornote{Corresponding author.}
\orcid{0000-0003-2990-1614}
\affiliation{%
  \institution{Beihang University, Beijing}
  \city{Yunnan Key Laboratory of Software Engineering}
  \country{China}
}
\email{lilicoding@ieee.org}


\begin{abstract}
  Deploying deep learning (DL) models on mobile applications (Apps) has become ever-more popular. However, existing studies show attackers can easily reverse-engineer mobile DL models in Apps to steal intellectual property or generate effective attacks. 
  A recent approach, Model Obfuscation, has been proposed to defend against such reverse engineering by obfuscating DL model representations, such as weights and computational graphs, without affecting model performance. 
  These existing model obfuscation methods use static methods to obfuscate the model representation, or they use half-dynamic methods but require users to restore the model information through additional input arguments. 
  However, these static methods or half-dynamic methods cannot provide enough protection for on-device DL models. Attackers can use dynamic analysis to mine the sensitive information in the inference codes as the correct model information and intermediate results must be recovered at runtime for static and half-dynamic obfuscation methods.
  We assess the vulnerability of the existing obfuscation strategies using an instrumentation method and tool, \InstrumentationToolname, that dynamically extracts correct sensitive model information (\ie weights, computational graph) at runtime. Experiments show it achieves very high attack performance (\eg 98.76\% of weights extraction rate and 99.89\% of obfuscating operator classification rate).
  To defend against such attacks based on dynamic instrumentation, we propose  \ObfuscationToolname, a Dynamic Model Obfuscation strategy similar to Homomorphic Encryption. The obfuscation and recovery process can be done through simple linear transformation for the weights of randomly coupled eligible operators, which is a fully dynamic obfuscation strategy.
  Experiments show that our proposed strategy can dramatically improve model security compared with the existing obfuscation strategies, with only negligible overheads for on-device models. Our prototype tool is publicly available at \href{https://github.com/zhoumingyi/DynaMO}{https://github.com/zhoumingyi/DynaMO}.
\end{abstract}

\maketitle

\section{Introduction}


More and more mobile applications (Apps) and IoT devices are leveraging deep learning (DL) capabilities.
Deploying DL models on such devices has gained great popularity as it avoids transmitting data and provides rapid on-device processing. It also enables applications to access their DL model offline.
As the computing power of mobile and edge devices keeps increasing, it also reduces the latency of model inference and enables the running of large on-device models.

However, as such DL models are directly hosted on devices, attackers can easily unpack the mobile Apps, identify DL models through keyword searching, extract key information from the DL models, formulate attacks on the models, or even copy them and reuse them in their own software.
This accessible model key information thus makes it easy to launch attacks or steal the model's intellectual property~\cite{zhou2024investigating}. 
To protect such on-device DL models, TFLite, the most commonly used on-device DL model framework, compiles the general DL model (such as TensorFlow and PyTorch models) to TFLite models, which disables direct white-box attacks. 
This is done by disabling the gradient calculation of the on-device models, which is essential for conducting effective white-box attacks. Such models are called non-differential models (\ie non-debuggable models).  
Such model compilation makes it hard for attackers to reverse engineer the on-device model. However, these on-device platforms still suffer from significant security risks. Recent attack methods~\cite{huang2022smart,li2021deeppayload,chen2022learning, zhou2024investigating} can parse model information in the on-device model (\eg \texttt{.tflite}) files and then reverse engineer them.

Recent model obfuscation approaches propose to use static or dynamic methods to obfuscate the representation of on-device models~\cite{zhou2023modelobfuscator, zhou2024model, mindspore}. 
Such DL model representations produced by model obfuscation methods cannot be understood by automatic tools or humans, but will not affect the model performance ~\etal \cite{zhou2023modelobfuscator}. The information inside model representations (\eg .tflite files) is protected by elaborating static obfuscation strategies, \eg operator renaming. However, DL models obfuscated by fully static methods must have the same computing process (\ie the whole data computing process from inputs to outputs) as the original model. The Mindspore platform~\cite{mindspore} uses a `half-dynamic' approach that produces an obfuscated model that defines different computing processes to the original model,  requiring additional input arguments from users to restore the correct computing process at the runtime. As developers need the correct input arguments to get the correct model output in mobile Apps, attackers can also find the arguments by unpacking the App and extracting the actual API calling steps~\cite{ren2024demistify}. This `half-dynamic' obfuscation can be considered a special form of static DL model obfuscation. 

To analyse and understand the limitations of these current static or half-dynamic obfuscation strategies, we designed a dynamic instrumentation method, \InstrumentationToolname. \InstrumentationToolname can identify the actual inference function of each operator, extract the sensitive data (\eg model weights), and identify the obfuscating components of the obfuscated DL model through dynamic instrumentation.
Our tool can recover nearly all obfuscated information to the original form (\eg 100\% of weights extraction rate and 99.87\% operator classification accuracy). This shows a major need to use robust obfuscations that can defend against dynamic instrumentation.
We then propose a fully dynamic obfuscation strategy, that we call Dynamic Model Obfuscation and build a tool \ObfuscationToolname to realiase it. \ObfuscationToolname adopts an idea similar to Homomorphic Encryption to randomly sample eligible operators to form obfuscation propagation paths to obfuscate the information from the start of the path and recover the results at the end of the path, thus building a random obfuscation-recovery operator pair. The process better secures the obfuscated model by the intermediate results and model information inside the obfuscation propagation path is obfuscated and will no be recover. The obfuscation and recovery process can be performed through simple linear transformation of the DL model weights, which avoids introducing overhead to the inference process and is hard for attackers to identify. 
Such dynamic DL model obfuscation can prevent instrumentation methods from collecting correct information about the inference code of each operator. Our experiments show that our method can significantly secure the model information compared with existing mobile DL model obfuscation strategies. Our proposed approach also introduces negligible overheads to the model inference.

The key contributions in this work include:

\begin{itemize}[leftmargin=*]
    \item We propose an attacking method using dynamic instrumentation to demonstrate key limitations with existing model obfuscation strategies. This can automatically acquire real information from the model inference functions at runtime.
    \item We analyse the limitations of existing obfuscation methods and propose a novel solution that can defend against the proposed instrumentation. It introduces obfuscation to the intermediate results of model inference.
    \item We have shown that our \ObfuscationToolname method can significantly increase the obfuscation in the model inference process with only negligible performance and efficiency loss compared with existing obfuscation strategies. 
    \item We open-sourced our prototype tool \ObfuscationToolname~\cite{zhou2024dynamo} in a GitHub repository:~\href{https://github.com/zhoumingyi/DynaMO}{https://github.com/zhoumingyi/DynaMO}.
\end{itemize}

In the following sections, we provide a motivation for our work and introduce the basic background of this study. We then conduct an experiment to demonstrate the key limitations of static and half-dynamic obfuscation methods. We propose a new approach and tool, and evaluate this on real-world mobile DL models. We discuss key findings, limitations and future research directions.


\section{Background and Related Works}

\subsection{Terminology}
According to their defending performance against different kinds of attacks, we use static method and dynamic methods to distinguish the different obfuscation methods. We call the existing model obfuscation methods~\cite{zhou2023modelobfuscator} as \textbf{static model obfuscation} as they only obfuscate the model representation in the compilation. So, they can just defend against the attacks based on static analysis for on-device models. In contrast, our method is \textbf{dynamic model obfuscation} as it can obfuscate the model information that is generated at runtime. Thus, it can defend against dynamic instrumentation. 

\subsection{DL Frameworks}

\paragraph{\textbf{Deep Learning (DL) Frameworks:}} The open-source community has developed many well-known \textbf{DL frameworks} to facilitate users to develop DL models, such as TensorFlow~\cite{tensorflow2015_whitepaper}, Keras~\cite{chollet2018keras}, and PyTorch~\cite{paszke2019pytorch}. These frameworks provide standards for developing DL models~\cite{dilhara2021understanding}. PyTorch is one of the latest DL frameworks which has gained academic user popularity for its easy-to-use and high performance. In contrast, TensorFlow is widely used by industry to develop new DL-based systems because it has the most commonly used on-device DL library, Tensor Flow Life (TFLite).  TFLite is the most popular library for DL models on smartphones, as it supports various hardware platforms and operation systems. 


\subsection{On-device DL Frameworks}

\paragraph{\textbf{On-device DL Frameworks:}} 
TensorFlow provides a tool \textit{TensorFlow Lite Converter}\footnote{\href{https://www.tensorflow.org/lite/convert/index}{https://www.tensorflow.org/lite/convert/index}} to convert TensorFlow models into TFLite models. A compiled TFLite model can then be run on mobile and edge devices. However, it does not provide APIs to access the gradient or intermediate outputs like other DL models. 

Traditionally, on-device models are released as \textbf{DL files} that are deployed on devices.
Mobile app code then accesses these models through a dedicated \textbf{DL library}, such as the TFLite library if the AI model is developed using the TFLite framework.
Each model file contains two types of information: \textbf{computational graph} and \textbf{weights}, which record the model's architecture and parameters tuned based on the training dataset, respectively. 
Such a computational graph is usually a multi-layer neural network.
In the network, each layer contains an \textbf{operator} that accepts \textbf{inputs} (i.e., the outputs of the previous operator), \textbf{weights} (i.e., stored in the dedicated file that is pre-calculated in the training phase), and \textbf{parameters} (i.e., configuration of the operator. For example, the \texttt{conv2d} layer in TFLite requires the parameters of stride size and padding type. Their parameters will affect the outputs of layers.) to conduct the neural computation and outputs the results for the next operator.

\subsection{DL Model Attacks}

DL models deployed on devices are subject to a range of attacks~\cite{papernot2017practical,zhou2020dast,li2021deeppayload,zhang2022investigating,huang2022smart,wu2024concealing}. These can include tricking the DL model with perturbed inputs into, e.g., classifying an image incorrectly; extracting model information to facilitate other attacks; stealing a copy of the model (which may have been very expensive to produce) for use in one's own application; and others. These attacks can be black-box~\cite{wu2020decision,li2021deeppayload,huang2022smart} or white-box~\cite{zhang2022investigating}. Access to DL components and/or access to DL models facilitates these attacks.

\subsection{Code Obfuscation}

Code obfuscation methods were initially developed to hide the functionality of malware. The software industry also uses it against reverse engineering attacks to protect code IP~\citep{schrittwieser2016protecting}. Code obfuscators provide complex obfuscating algorithms for programs like JAVA code~\citep{collberg1997taxonomy,collberg1998manufacturing}, including robust methods for high-level languages~\citep{wang2001security} and machine code level~\citep{wroblewski2002general} obfuscation. Code obfuscation is a well-developed technique to secure the source code. However, solely relying on traditional code obfuscation approaches cannot effectively protect on-device models, especially in terms of protecting the structure of DL models and their parameters. 

\subsection{Model Obfuscation}
\label{subsec:model_obfuscation}


To prevent attackers from obtaining detailed information on deployed DL models, model obfuscation has been proposed. This obfuscates model representations such as the weights and model architectures~\cite{zhou2023modelobfuscator}. The \textit{renaming} and \textit{parameter encapsulation} can prevent most model parsing or reverse engineering methods from extracting the key information (\eg weights and computation graph) of the deployed model. In the scenarios where computational costs are critical, the \textit{neural structure obfuscation} and \textit{shortcut injection} do not introduce any additional overhead as they just add misleading information to the model representation but will not modify any inference process. For structure obfuscation, developers can use \textit{shortcut injection} and \textit{extra layer injection}. These methods can increase the difficulty of understanding the model structure of the deployed models. 

The most vulnerable obfuscation methods are \textit{neural structure obfuscation} and \textit{shortcut injection}. These do not change any inference process but just use to mislead the attacker when parsing the model information. For \textit{neural structure obfuscation}, attackers can analyse the real data flow in the model inference process to obtain the real neural architecture from the shape of intermediate data. TFLite actually provides an official API to get such information. For \textit{shortcut injection}, we can slightly modify the model like the paper in~\cite{li2021deeppayload} to remove each shortcut and check whether the model inference works well. We can identify the obfuscating shortcut that will not actually be used in the inference. However, we cannot do the same process for \textit{extra layer injection} because the obfuscating extra layer actually participates in the model inference (although it will not affect the model output). Therefore, in our study, we mainly focus on analysing the robustness of three obfuscation methods (shown in Figure~\ref{fig:motivation}): \ie \textit{renaming}, \textit{parameter encapsulation}, and \textit{extra layer injection}.

\subsection{Motivation for Our Work}

\begin{figure}[!th]
  \begin{center}
    \includegraphics[width=1.0\linewidth]{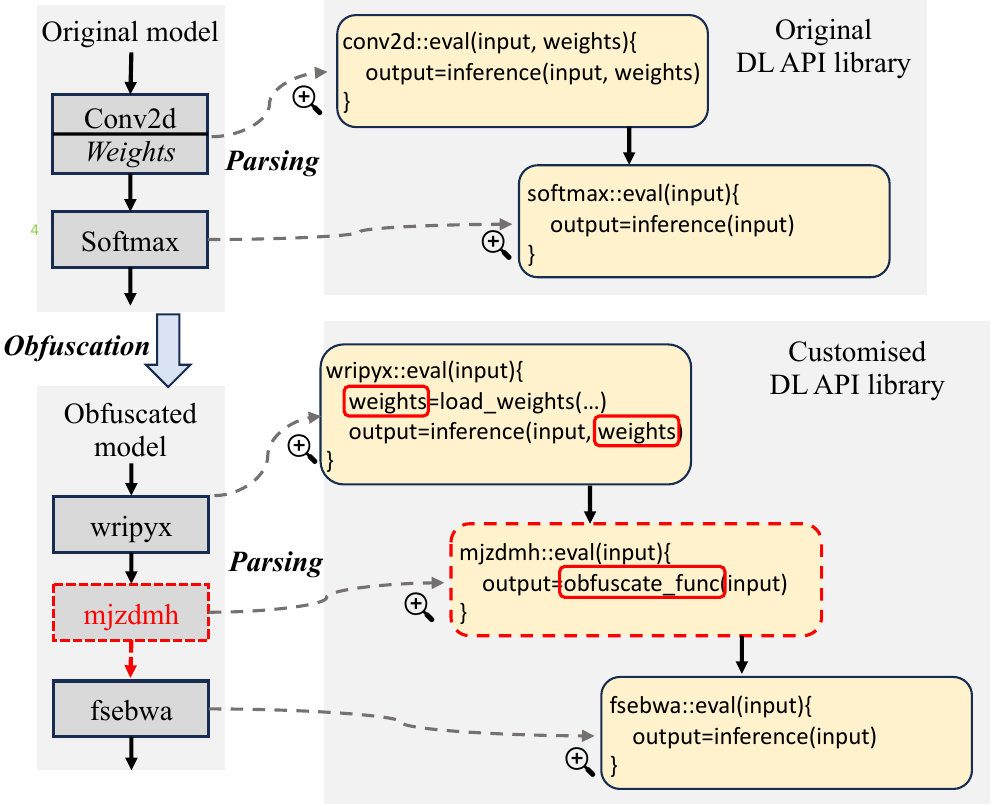}
  \end{center}
  \vspace{-1em}
  \caption{Demonstration of existing model obfuscations~\cite{zhou2023modelobfuscator}. Here, existing model obfuscation hides the weights of \texttt{conv2d} operator, renames the original operator name to random strings (\texttt{conv2d} $\rightarrow$ \texttt{wripyx}), and injects an extra obfuscating operator (\ie \texttt{mjzdmh}). The customised DL API library is generated to execute the inference of the obfuscated model. The function \texttt{\{OP\_NAME\}::eval} is the code implementation of the operator's forward inference. The extra operator \texttt{mjzdmh} only has an obfuscating function \texttt{obfuscate\_func} to copy the input value to the output. }
  \label{fig:motivation}
\end{figure}

Statically obfuscated DL model representations on mobile devices are still directly exposed to threats. As shown in Figure~\ref{fig:motivation}, an original DL API library will use the operator's name to locate the code implementation of the operator's forward inference to build a correct function call graph. We refer to this as the \textbf{\emph{inference code of DL operators}} in this paper.  Although the name of the obfuscated DL operators is randomly generated by  \textit{renaming}, \eg \texttt{conv2d} $\rightarrow$ \texttt{wripyx}, attackers can still use the operator's name to locate the inference code for each operator. For example, when the DL library gets an operator's name \texttt{\{OP\_NAME\}}, it will use the function \texttt{\{OP\_NAME\}::eval} to perform the forward inference of the operator.

Existing model obfuscation strategies use static or half-dynamic ways to obfuscate model representation. To produce correct model outputs, each obfuscated information or its produced results needs to be recovered in the inference code of operators at runtime.
\textbf{Therefore, attackers can use dynamic analysis to extract the correct model representation from the DL model's inference code at runtime.} For example, in Figure~\ref{fig:motivation}, attackers can hook a data collection function to the \texttt{wripyx::eval} to obtain the correct model weights. In addition, they can modify the \texttt{obfuscate\_func} function at runtime to identify whether an operator is an extra obfuscating operator (the extra operator should not affect the model output~\cite{zhou2023modelobfuscator}).

\subsection{Research Questions}

To assess and enhance the robustness of existing mobile DL model protections, this study aims to address the following key research questions:

\begin{itemize}[leftmargin=*]
    \item \textbf{RQ1 - What are the limitations of existing model obfuscation methods?}
    \item \textbf{RQ2 - How can we better defend DL models against dynamic instrumentation attacks?}
    \item \textbf{RQ3 - How efficient is our proposed obfuscation strategy?}
\end{itemize}


\section{RQ1: Model Deobfuscation}
\label{sec:deobf}
\begin{figure*}[t]
  \begin{center}
    \includegraphics[width=0.99\linewidth]{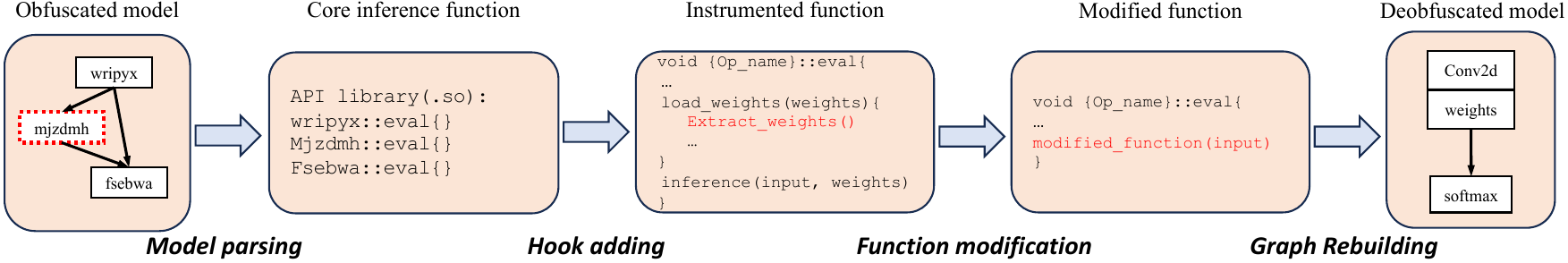}
  \end{center}
  \vspace{-1em}
  \caption{Overview of our proposed model deobfuscation method \InstrumentationToolname. The \texttt{mjzdmh} operator with a red dotted block is an extra obfuscating operator.}
  \label{fig:attack}
\end{figure*}


In this study, we use the most commonly used on-device model on Android, TFLite models, as our target.

\subsection{Threat Model}
The on-device DL models and their corresponding third-party API library used for model inference are packed into mobile applications on devices like Android. In the model deobfuscation process, attackers need to unpack the mobile application using reverse engineering tools like \textit{apktool}~\cite{apktool} to get the on-device DL models and related API libraries. The DL model representation is usually stored in a separate model file such as \texttt{.tflite} files for TFLite models. As the existing commonly used DL platforms are usually open-sourced like TensorFlow, PyTorch, and ONNX, even if the on-device model has been obfuscated, attackers can obtain the name of DL operators, such as a random string, used in the model inference for parsing the computational graphs. Thus, attackers can use the collected names of each operator to identify the actual inference function of each operator in the API library, because DL libraries use the operator name to determine the actual inference function at runtime. 

\begin{lstlisting}[caption={Simplified backend inference function of \texttt{Conv2d} operator for running on a multithreaded CPU.},captionpos=b]
void {op_name}::Eval {
  // Create essential data for the operator
  TfLiteTensor* input = create_input(input_data);
  TfLiteTensor* weights = create_weights(weights_data);
  ConvParams params = create_params(paras_data);
  TfLiteTensor* output;
  // The computing function for operator
  multithreaded::Conv(params, input, weights, output);
}
\end{lstlisting}

After identifying the inference function of each DL operator, attackers do not need to understand all the compiled code used for the operator inference in the DL API library, which is a very complex task. But as the source code of TFLite and its related tools are open-source, for example, they can identify the API of loading the \texttt{weights\_data} into the memory and be used to build the corresponding TFLite tensor (\ie \texttt{TfLiteTensor* weights} in Listing 1). Note that the construction processes of \texttt{weights} are the same in different DL operators. Thus, attackers can synthesise test samples and use dynamic instrumentation analysis to hook a data collection function in the inference code to get the real model information.

\subsection{\InstrumentationToolname}

Existing model obfuscation methods can obfuscate multiple kinds of model information such as the functionality of operators (\ie operator name), model weights, neural architectures, and the spatial relationship among operators. It is hard for attackers to parse the model representation without any auxiliary information. 
However, the model representations must store the names of the operators to provide the information on which inference codes need to be used to produce the model outputs. Attackers can use the operator's name (even if it is obfuscated) to identify the inference code of operators, thus attackers can use dynamic instrumentation and analysis to extract the correct model information inside the inference code. 


\paragraph{\textbf{Overview}} The overview of our proposed method \InstrumentationToolname is shown in Figure~\ref{fig:attack}. (1) Our method \InstrumentationToolname first parses the obfuscated model and determines which inference function will be used at runtime for each operator. (2) Then, our proposed tool can use the instrumentation method to hook a data collection function to the construction process of model weight data. This function can automatically extract the weights of each operator and save them into a separate file. (3) Next, our tool will automatically modify the actual inference function for each operator to filter out which operator is the obfuscating operator. The obfuscating operator produces the output that is equal to the input or modifies the output but will not change the output of the next operator. (4) Finally, after getting the weights and the real computation graph of the on-device models. Our tool will then analyse the relation between inputs, outputs, and weights to infer the functionality (\ie the real name) of operators. We implement \InstrumentationToolname using dynamic program instrumentation framework \textbf{\textit{Pin}}~\cite{pin}.

\paragraph{\textbf{Model Parsing}} We first write a script to automatically synthesise the inputs as the specification and feed them into the on-device model. Thus we can use dynamic instrumentation and analysis to parse the model. We need to collect the names of operators (obfuscated as random strings) in the DL model. Our tool traces the execution process of the model inference, and then identifies the executed API functions that contain the operator's name. In the implementation of TFLite, each operator's source code will have four basic functions: \texttt{prepare} (pre-allocate the intermediate data), \texttt{init} (initialize the essential data), \texttt{free} (remove the data after the inference), \texttt{eval} (the computing code of inference). The source codes of each function will be written in a C++ \texttt{workspace} named as the operator name. Therefore, we can use the operator's name in model representations to identify the corresponding workspace and locate the \texttt{eval} function which has the code implementation of model inference. The identified inference code will be used in the dynamic instrumentation to deobfuscate the model information.

\paragraph{\textbf{Hook Adding}}

After identifying the core inference function of each operator, \InstrumentationToolname extracts the model weights of each operator. Although the weights can be obfuscated (\eg store them in different places and assemble them before inference), the inference function needs the correct value to perform the right DL model inference computing process. Therefore, \InstrumentationToolname will search for the weight tensor construction function in the inference function of the DL API library using dynamic analysis, to find the place where the weights data is loaded into the memory and can be extracted. As the TFLite and its related tools are open-sourced, we can get the keyword of the weight tensor construction function. \InstrumentationToolname will identify the tensor construction step of the API library through keyword searching. If the related construction function starts to be executed in the inference process of operators, our \InstrumentationToolname will automatically hook a data extraction function into the start of the construction to parse the input data of the construction function and store the extracted weights data in a separate file. Note that, \InstrumentationToolname will also extract the parameter of the operator if it has, such as the padding value and stride size for \texttt{conv2d} operator. 
The parameter is stored as a \texttt{struc} data \texttt{params} in C++ source codes of TFLite. 
The process of extracting parameters is similar to model weights, so we omit the implementation details here. Then, the instrumented API library will automatically extract and save the real weights used in the model inference at runtime.

\begin{table*}[t]
\small
\centering
\caption{Deobfuscation performance of the proposed \InstrumentationToolname. WER, WEA, OCA, NIR, and SS are the five metrics used to measure deobfuscation performance.}
\begin{tabular}{lcccccccccc|c}
\hline
 Metric   & Fruit & Skin &MobileNet&MNASNet &SqueezeNet &EfficientNet &MiDaS &LeNet &PoseNet &SSD & \textbf{Average value} \\
\hline
WER      & 100\% & 100\% & 100.0\%     & 98.15\%   & 96.30\%      & 98.04\%    &  95.10\% &  100.0\%       & 100\%    & 100\%  & \textbf{98.76\%}        \\
NIR  & 100\% & 100\% & 96.77\%     & 98.41\%   & 97.50\%      & 100\%    &  91.18\% &  100\%       & 100\%    &  100.0\%     & \textbf{98.39\%}        \\
OCA  & 100\% & 100\% & 100\%     & 98.88\%   & 100\%      & 100\%    &  100\% &  100\%       & 100\%    &   100\%    & \textbf{99.89\%}        \\
\hline
WEE      & $2.8 \times 10^{-5}$ & $1.3 \times 10^{-5}$ & $2.6 \times 10^{-5}$     &$3.0 \times 10^{-5}$   & $3.6 \times 10^{-5}$      & $2.3 \times 10^{-6}$    &  $1.9 \times 10^{-5}$ &  $8.8 \times 10^{-7}$       & $2.5 \times 10^{-5}$     & $2.5 \times 10^{-5}$  & $\bf{2.1 \times 10^{-5}}$       \\
SS  & 1.0 & 1.0 & 0.97     & 0.98   & 0.97      & 1.0    & 0.91 &  1.0      & 1.0    &   1.0    & \textbf{0.98}        \\
\hline
\end{tabular}
\label{tb:deobf}
\end{table*}

\paragraph{\textbf{Function Modification}}
Another key aspect of on-device DL models is their computational graph. This can be obfuscated by extra obfuscating operator injection. Such extra obfuscating operators are used to participate in the inference process but will not affect the final results. For example, they can just produce the same output as the input, or produce an output out of the range of the input shape of the next operator which will not affect the results of the next operator. In the function modification process, \InstrumentationToolname will first monitor the inference process at runtime. As shown in Listing 2, if a function \texttt{\{op\_name\}::eval()} starts to execute, our \InstrumentationToolname will dynamically insert a value-copy function to the inference function, \ie the \texttt{eval()} function, and add a stop signal of the inference function. For instance, the TFLite inference function will finally \texttt{return} a \texttt{TFLite\_OK} (\ie equal to 0) to stop the execution and jump to the inference process of the next operator. We use the value-copy function as the modified computing function which will produce the same output as the input. Next, \InstrumentationToolname will compute the output difference between the instrumented API library and the original library for each operator under the same input. If the two outputs are not identical, it means the modified operator is a valid operator that belongs to the original model. Otherwise, the modified operator is an extra obfuscating operator because the modified function will not affect the model outputs.

\begin{lstlisting}[caption={Modified inference function of DL operator},captionpos=b]
void {op_name}::Eval {
  // Create essential data for the operator
  TfLiteTensor* input = create_input(input_data);
  TfLiteTensor* output;
  // Copy the value of input to output
  copy_value(input, output);
  return TFLite_OK;
  // The original implementation of the operator
  ...
}
\end{lstlisting}

\paragraph{\textbf{Graph Rebuilding}}
After removing the extra obfuscating operators and extracting the real model weights for each operator, the only obfuscated information we left is the functionality of the operator, \ie the name of the operators. For the operators with weights, the functionality of operators can be inferred by the data shape (\ie input, output, and weights). For example, the output shape of \texttt{conv2d} operator can be found as follows:

\begin{equation}
\label{eq:cond_size}
\begin{aligned}
H_{\text{out}} & =\left\lfloor\frac{H_{\text{in}}+2 \times \text{P}[0]-\text{D}[0] \times ( \text{W\_Size}[0]-1)-1}{\text{stride}[0]}+1\right\rfloor \\
W_{\text{out}} & =\left\lfloor\frac{W_{\text{in}}+2 \times \text{P}[1]-\text{D}[1] \times ( \text{W\_Size}[1]-1)-1}{\text{stride}[1]}+1\right\rfloor
\end{aligned}
\end{equation}

Where the $H_{in}$ and $H_{out}$ are the dimensions in height of input data and output data, respectively. The $P$ (\ie padding), $D$ (\ie dilation), and stride are the parameters (\ie the setting) of the \texttt{conv2d} operator. $W\_Size$ is the shape of the weights. In commonly used DL models, there are only a limited number of operators with weights like \texttt{conv2d}, \texttt{depthwise\_conv2d}, \texttt{fullyconnected}. These operators have different output shapes for the same input and weights shape. In our \InstrumentationToolname, it will use the output size to identify the name of operators according to the predefined size transformation rules of potential operators (\texttt{conv2d}, \texttt{depthwise\_conv2d}, and \texttt{fullyconnected}), \eg Equation~(\ref{eq:cond_size}). 
For the operators without weights (\eg \texttt{softmax}, \texttt{relu}, and \texttt{add}), \InstrumentationToolname will extract the input and output data of each obfuscated operator at runtime. Then, it uses the input to test different operators in a pre-collected list. Note that to produce this operator list, we can collect the operators from the TFLite operator list. Because the parameter data of these operators have various construction processes, we cannot extract the parameter data for all operators. So, to develop a robust analysing method, we do not collect the parameter data to predict the functionality of operators without weights. We use the input-output relationship to guess the parameter of each candidate operator and compute the output using the guessed parameter. If the test operator produces the same output as the obfuscated operator of the on-device model, the name of the test operator is the name of the obfuscated operator as they share the same functionality. Thus, our proposed \InstrumentationToolname can parse the functionality of each valid operator with a random obfuscating name.

\subsection{Deobfuscation performance}


\subsubsection{\textbf{Dataset}} To evaluate \InstrumentationToolname's performance on deobfuscating on-device DL models with various structures for multiple tasks, we use the 10 TFLite models, as used in \cite{zhou2023modelobfuscator}. These include a fruit recognition model, a skin cancer diagnosis model, MobileNet~\citep{howard2017mobilenets}, MNASNet~\citep{tan2019mnasnet}, SqueezeNet~\citep{iandola2016squeezenet}, EfficientNet~\citep{tan2019efficientnet}, MiDaS~\citep{ranftl2020towards}, LeNet~\cite{lecun1998gradient}, PoseNet~\citep{kendall2015posenet}, and SSD~\citep{liu2016ssd}. 
The fruit recognition and skin cancer diagnosis models are collected from Android apps. The other models were collected from the TensorFlow Hub~\footnote{\url{https://tfhub.dev/}}. 

\subsubsection{\textbf{Results}}

We first evaluate the deobfuscation performance of \InstrumentationToolname. Existing model obfuscation methods can obfuscate multiple model information that includes weights, the functionality of each operator, and computational graphs. \textbf{We use five deobfuscation metrics to measure the performance of \InstrumentationToolname, including Weights Extraction Rate (WER), Weights Extraction Error (WEE), Operator Classification Accuracy (OCA), Name Identification Rate (NIR), and Structure Similarity (SS).} Note that the two metrics WEE and SS represent the global performance in weight and structure extraction.

The formula of WER is $\textbf{WER} = n/m$, where $n$ is the number of extracted weights with an element-wise maximal error of the output difference less than $1\times10^{-4}$ compared with the original weights. The original weights are extracted by official APIs provided by TFLite. We choose $1\times10^{-4}$ as the threshold because we consider the error in computing and extraction (The error in extracting the original weights of the model using TFLite APIs is within 5 decimal places.). $m$ is the total number of weights that need to be extracted. The WEE is used to measure the average difference of the extracted weights and original weights, which can be formularized by $\textbf{WEE} = \frac{1}{N}\sum_1^N\max(|W'_n - W_n|) $, where $W'_n$ and $W_n$ are the extracted weights and the original weights of the $n-$th operator, respectively. Note that we only compute the WEE for the successfully extracted weights (the maximal error of the output difference less than $1\times10^{-4}$ compared with the original weights). The \textbf{OCA} is a binary classification metric. We only classify whether an operator is a valid operator or an obfuscating extra operator. For NIR, it can be computed by $\textbf{NIR} = i/j$, where $i$ is the number of successfully identified operators, and $j$ is the total number of operators that have been classified as valid operators. We collect an operator list (See our code repository), and identify which operator in the list the renamed operator is equal to. For SS, if one operator is misclassified in NIR or OCA, it can be considered an error point. Then we use the number of error points divided by the number of operators to compute the SS. 

The deobfuscation results are shown in Table~\ref{tb:deobf}. The \InstrumentationToolname achieves 98.76\% of Weights Extraction Rate (WER), 99.89\% of Operator Classification Accuracy (OCA), 98.39\% of Name Identification Rate (NIR), and 0.98 pf Structure Similarity (SS). That means our instrumentation analysis method can effectively deobfuscate the on-device model. In addition, our method achieves low Weights Extraction Error (WEE).

\subsubsection{\textbf{Analysis}}

Our results show that existing mobile DL model obfuscation methods are not robust. The reason is we need to deploy the model representation on mobile devices to guarantee the DL API library performs the right inference computations. Existing model obfuscation strategies need to restore the correct information and execute the correct inference process for each operator at runtime, thus enabling the attacking exploitation based on the dynamic instrumentation. In addition, \InstrumentationToolname only needs a maximum of $2\times n$ (some operators don't have any weights) times of inference to collect the information of an n-layer model. In our experiments, attackers only need 5-20 minutes to finish all steps for one model. Therefore, existing obfuscation methods are very vulnerable to dynamic instrumentation-based attacks.
This shows that we need to develop a new approach to defence against attacks based on dynamic instrumentation.

\begin{tcolorbox}[colback=gray!5!white,colframe=gray!85]
RQ1: Our \InstrumentationToolname can effectively extract the sensitive information of obfuscated models generated by existing obfuscation strategies. Existing obfuscation methods are not robust for the instrumentation-based attack.
\end{tcolorbox}

\begin{figure*}[t]
  \begin{center}
    \includegraphics[width=0.96\linewidth]{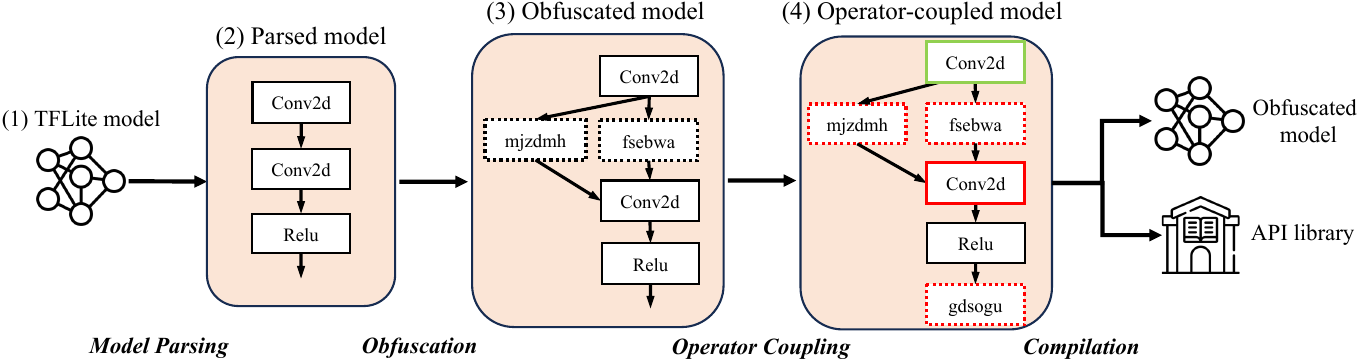}
  \end{center}
  \vspace{-1em}
  \caption{Overview of our proposed \ObfuscationToolname. The dotted block denotes the extra obfuscating operator. The green block denotes the selected operator in weight transformation obfuscation. The red block (both dotted and solid) denotes the eligible coupled operators that can be used to recover the correct results. Note that we do not obfuscate the names of valid operators to demonstrate the process clearly.}
  \label{fig:mod}
\end{figure*}

\section{RQ2: Dynamic Obfuscation Strategy}

\subsection{Approach Requirements}

Despite using  \textit{renaming}, \textit{parameter encapsulation}, and \textit{extra layer injection} static obfuscation methods, RQ1 results show that attackers can use dynamic instrumentation to extract correct DL model information at runtime.  
Although there are many existing methods, such as various cryptographic techniques, that might be used to protect a DL model, it is hard to use them to protect on-device models, as such a model is very sensitive to computational cost. In addition, we should avoid requiring additional hardware support, as existing mobile platforms like Android cannot provide specific needed hardware/software environments to millions of Apps. 

To disable instrumentation-based analysis, as used in \InstrumentationToolname, we need to not only provide the obfuscating information to the attackers but also feed obfuscating information to the inference code of operators. However, it is hard to feed the obfuscating information to the inference process but finally restore the results to the correct one, as well as add enough randomness to the process. For example, a straightforward strategy is transforming the weights or outputs of one operator and performing the inverse process in the next operator (similar protection has been used in existing DL platforms like~\cite{mindspore}). Such an obfuscation can be easily identified because it requires the next operator to perform a reverse transformation after transforming the weights in the previous operator (it has a different inference pattern than the normal DL model). Attackers can also use dynamic instrumentation to get the details of the result recovery process in the next operator and then deobfuscate the weights of the operator.
If the weights of a valid operator $i$ (the inference formula is $X_{i+1} = f(W_{i}^TX_{i})$) are obfuscated by $aW_i$, the correct results of the operator need to be recovered by performing $\frac{1}{a}X_{i+1}$ in the start of the next operator. Attackers can identify the result-recovering process as the inference code needs to be modified to recover the correct results (adding another step $\frac{1}{a}X_{i+1}$). Then, they can use dynamic instrumentation to obtain the recovering parameter $\frac{1}{a}$, and deobfuscate the weights of the operator $\frac{1}{a}(aW_i) = W_i$.


\subsection{The Dynamic Model Obfuscation Process}

Our solution to defend against dynamic instrumentation adopts a concept akin to Homomorphic Encryption -- that is,  we use obfuscated information (\eg weights, architecture) to perform DL model inference to get \emph{obfuscated results}. These results can then be recovered by a \emph{deobfuscation step} after the computing process of several operators. 
To do this, we perform simple linear transformations for model weights in randomly picked operators. 
Then, we randomly choose eligible operators (including the extra obfuscating operators), which may not be the directly connected next operator. The extra obfuscating operators can be considered a special linear operator whose weight data is an identity matrix. It injects the corresponding inverse transformation in the weights of chosen eligible operators to recover the model results. This approach prevents attackers from using dynamic instrumentation in actual inference code functions of each operator to obtain the sensitive model information because the relation between the representation obfuscation and information recovery at runtime does not have a fixed pattern.
It then becomes very hard to reverse engineer the steps of weight transformation and corresponding result recovery, even if attackers can parse one of these. In addition, extra obfuscating operators can be added to contribute to the model output, \ie recover the correct intermediate results using a linear transformation like other normal operators. Attackers cannot use dynamic code modification to identify such extra obfuscating operators. 

The key steps of our proposed method, Dynamic Model Obfuscation (\ObfuscationToolname), are outlined in Figure~\ref{fig:mod}. We integrate our obfuscation strategy into the existing model obfuscation process. Our tool \ObfuscationToolname first parses the on-device TFLite model. It then obfuscates it using existing model obfuscation methods, producing obfuscated operators. Compared with the existing obfuscation process, our method has one more step for coupling these obfuscated DL model operators. It uses a fully dynamic obfuscation method to compute intermediate results using obfuscated weights from a selected operator, and then recovers the correct results at the coupled operator.
To achieve this, \ObfuscationToolname obfuscates the weights of the selected operator like existing methods (\ie using simple linear transformation), but recovers the correct results by another weight transformation. That means \ObfuscationToolname does not need to add an additional step to recover the results which is easy to identify. Attackers cannot identify it unless they know the original weights.

Unlike existing strategies, the choice of coupled operators has multiple possibilities. For example, it can obfuscate the weights of the green \texttt{conv2d} operator in Figure~\ref{fig:mod}, and choose the \texttt{mjzdmh} and \texttt{fsebwa} or following \texttt{conv2d} operators as the coupled operators. This can further increase the difficulty for attackers to identify the {operator pair (\ie selected operator and coupled operator)}, even if attackers manage to identify one of them. In addition, previous extra obfuscating operator injection methods can only inject obfuscating operators that will not change the final output to the computation graph. In our \ObfuscationToolname, the extra obfuscating operators (\eg \texttt{mjzdmh} and \texttt{fsebwa}) perform a simple linear computation, and can affect the final output when they are chosen as the selected operator or the coupled operator without significantly increasing the computational overhead. Thus, it is hard for the dynamic instrumentation and analysis proposed in \InstrumentationToolname to identify the extra obfuscating operator as it performs a similar linear computation process to other operators.

In summary, we propose a fully dynamic obfuscation strategy. It can obfuscate the intermediate results and model information without performance loss. It also increases the randomness in the obfuscation process, including the choice of the selected operator and the coupled operator, with negligible overhead compared with existing obfuscation methods. Thus, \ObfuscationToolname can significantly increase the difficulty of the model deobfuscation. Note that we do not need to obfuscate all weights of on-device models if attackers cannot identify which weight data is obfuscated.

\subsection{Obfuscation}

The first step is obfuscating the on-device DL model using the existing obfuscation methods, including \textit{Renaming}, \textit{Weights encapsulation}, \textit{Neural architecture obfuscation}, \textit{Extra layer injection}, and \textit{Shortcut injection}. The \textit{Neural architecture obfuscation} and \textit{Shortcut injection} are compatible with our strategy, but they will not affect the outcome of our method whether they exist in the obfuscation process or not. Therefore, we omit them in our analysis and visualization (\ie Figure~\ref{fig:mod}). After parsing and obfuscating the on-device model using an existing tool like ModelObfuscator~\cite{zhou2023modelobfuscator}, we obtain the obfuscated DL model,  shown in Figure~\ref{fig:mod}.

\subsection{Obfuscation Coupling}

\textbf{Notations:} \ObfuscationToolname will then perform obfuscation coupling to introduce the obfuscation to the intermediate results of the computing process. As we mentioned in the Overview section, consider a linear operator, we can obfuscate the weight of the selected operator using a linear transformation, which can be formulated as:
\begin{align}
\label{eq:linear_trans}
    X_{n+1} = aW_{n}^TX_{n} + ab_{n}
\end{align}
where $X_{n}$ is the input of the n-th operator of the DL model, $x_{n+1}$ is the output of the $n$-th operator and also the input of $(n+1)$-th operator. $W_{n}, b_{0}$ are the weights of the operator. In this linear transformation, we use the product of a scale value $a$ and original weights $W_{n}, b_{0}$ as the new weights. We can use such linear transformation to obfuscate the weights of any linear operators, including \texttt{FullyConnected}, \texttt{Conv2D}. The formula for \texttt{Conv2D} is different from the Equation~\ref{eq:linear_trans}, but it has similar properties. So, we use it to present all linear operators. \textbf{Note that, the extra injecting operator in existing model obfuscations also can be represented as a linear operator, where the $W_n$ is equal to an identity matrix $I_n$, and the $b_n$ is a zero vector.} Thus, the weights of the selected operator can be obfuscated. Next, we need to find the coupled operator to perform another linear transformation to the weights to recover the correct results. To show the process clearly, we first introduce the Coupled Weight Transformation rule for linear operators, which is shown as follows:
\begin{lemma}
\label{lemma_1}
     \textbf{(Coupled Weight Transformation on linear model)}: A sub-network $f$ consists of multiple linear layers $\{L_1, L_2, \cdots, L_n\}$ and the $i$-th layer is $L_i: X_{i} = W_{i-1}^\top X_{i-1} + b_{i-1}$ where $i\in [1, n]$. The output of the sub-network $f$ \wrt to the input $X_0$ would be $f(X_0)$.
    If $W_1$ is transformed to $aW_1$, $W_{n}$ is transformed to $\frac{1}{a}W_{n}$, and $b_i$ is transformed to $ab_i$ for $i\in [1, n-1]$, then the transformed network $f^s$ \wrt to the input $X_0$ would be $f^s(X_0)$ and we have $f^s(X_0) = f(X_0)$. \textbf{The proof can be found in our~\href{https://github.com/zhoumingyi/DynaMO/blob/main/ASE2024_Sup.pdf}{code repository}.}
\end{lemma}

\noindent \textbf{Explanation of Lemma~\ref{lemma_1}}: Through the Lemma~\ref{lemma_1}, if we obfuscate the weights by a linear transformation (\ie $aW_{0}$, $ab_{0}$) in the first \texttt{conv2d} (the selected operator) with a green box (shown in Figure~\ref{fig:mod} (4)), We can recover the corrected results by performing a linear transformation (\ie $\frac{1}{a}W_{n}$, $\frac{1}{a}b_{n}$, where $n$ denotes the operator ID of the coupled operator) in the coupled operator. The coupled operator could be \{\texttt{mjzdmh}, \texttt{fsebwa}\} as the output of the selected operator (the green \texttt{conv2d}) are shared to both \texttt{mjzdmh} and \texttt{fsebwa}, or the coupled operator could be the second \texttt{conv2d} with a red box. The coupled operator can not be \texttt{gdsogu} in Figure~\ref{fig:mod}, as the previous operator \texttt{relu} is not a linear operator. So the coupled Weight Transformation rule will not exist for it.

We can follow the Coupled Weight Transformation rule to search the operator pair on linear models. Although the DL model usually has non-linear operators like \texttt{relu}, we can still find many eligible coupled operator pairs for the non-linear model unless all linear operators are followed by a non-linear operator. However, the most commonly used on-device DL architecture Convolutional Neural Networks (CNNs) usually have many nonlinear operators, \ie some \texttt{conv2d} operators followed by a \texttt{relu} operator in the Convolutional Neural Network, and the extra obfuscating operator cannot be injected between the \textit{conv2d} and the followed \texttt{relu} operator as they are fused as one operator in TFLite to increase the inference efficiency. Thus, searching for the potential coupled operator will usually fail if we choose the \textit{conv2d} as the selected operator to obfuscate. Therefore, to increase the utility of our method and extend the Coupled Weight Transformation rule for the propagation path has the nonlinear operator, we define a Coupled Weights Obfuscation rule for non-linear models, which is shown as follows:

\begin{theorem}
\label{theorem_1}
      \textbf{(Coupled Weights Obfuscation)}: We consider a general non-linear $\text{ReLU}_\beta$ layer (\eg \texttt{relu6} operator):
      \begin{align}
        \label{eqn:relu_a}
        \text{ReLU}_\beta (x) =
        \begin{cases}
          \beta, &\text{if} \quad x \geq \beta; \\
          x,      &\text{else if} \quad 0 < x < \beta; \\
          0,      &\text{otherwise}. \notag
        \end{cases}
      \end{align}
      If $W_i$ and $b_i$ in the sub-network $f$ are scaled to $aW_i$ and $ab_i$ for $i\in [1, n]$ with $0<a<1$, respectively, to get the transformed sub-network $f^s$. Then, $\text{ReLU}_\beta ( \frac{1}{a}I_{n+1} \times \text{ReLU}_\beta \big(f^s(X_0)\big) ) = I_{n+1} \times \text{ReLU}_\beta \big(f(X_0)\big) $. \textbf{The proof can be found in our~\href{https://github.com/zhoumingyi/DynaMO/blob/main/ASE2024_Sup.pdf}{code repository}.}

\end{theorem}

\noindent \textbf{Explanation of Theorem~\ref{theorem_1}}: When we choose the first \texttt{conv2d} operator with a green box (see Figure~\ref{fig:mod}), the transformation propagation can be extended to the next extra obfuscating operator connected to the first \texttt{relu} in the transformation path, \ie \texttt{gdsogu} operator that performs $\text{ReLU}_\beta (\frac{1}{a}I_{n+1} \times \text{ReLU}_\beta \big(f^s(X_0)\big))$, where the $I_{n+1}$ is the weights of the operator and is an identity matrix. Specifically, \ObfuscationToolname will perform a linear transformation to the weights ($aW_{0}$, $ab_{0}$) as it is the selected operator. Then, except for choosing \{\texttt{mjzdmh}, \texttt{fsebwa}\} or \texttt{conv2d} as the coupled operator (as shown in the explanation of Lemma~\ref{lemma_1}), we can choose the \texttt{gdsogu} operator as the coupled operator by transforming the weights (\ie $\frac{1}{a}W_{4}$ and $\frac{1}{a}b_{4}$, where $W_{4}$ is an identity matrix and $b_{4}$ is a zero vector). Note that here we need to add a fused \texttt{relu} computation at the end of the inference process of \texttt{gdsogu} (\ie the outer \texttt{ReLU} function of $\text{ReLU}_\beta ( \frac{1}{a}I_{n+1} \times \text{ReLU}_\beta \big(f^s(X_0)\big) )$).
By adding many extra obfuscating operators to the computational graph, \ObfuscationToolname can find more eligible transformation pairs for CNNs, thus improving the obfuscation performance. 

\begin{table}[t]
	\centering
        \small
	\begin{tabular}{lll}
		\toprule
		$\bf Algorithm~1:$ Obfuscation Coupling   \\
		\midrule

		\textbf{Input}:$ \text{ computational graph } \mathcal{G}, \text{ the number of obfuscation pairs } n.$  \\
		\textbf{Output}:$ \text{ obfuscation-coupled graph } \mathcal{G} $\\
            $ 1: \text{Initialize a coupled operators set } \textbf{S}, \text{ and a transformation parameter } $ \\
            $ \ \ \ \ \ \text{set } \textbf{A} $ \\
		$ 2: \textbf{For } \ m \text{ in range} (1, n)  \textbf{ do}:$ \\
            $ 3: \qquad \text{random choose an eligible selected operator } O  \text{ in } \mathcal{G}$ \\
		$ 4: \qquad \text{find the eligible coupled operator set } \overline{\textbf{O}} $ \\
		$ 5: \qquad \textbf{if } \overline{\textbf{O}} \text{ is not empty:}$ \\
		$ 6: \qquad \qquad \text{random choose the coupled operator } \overline{O} \text{ from } \overline{\textbf{O}}$  \\
  		$ 7: \qquad \qquad \textbf{S.append}(O,\overline{O}) $  \\
            $ 8: \qquad \qquad  a = \textbf{random}(0,1)$  \\ 
            $ 9: \qquad \qquad \textbf{A.append}(a) $  \\
            $10: \textbf{For }  (O,\overline{O}) \text{ in } \textbf{S}: $ \\        
            $11: \qquad \text{perform transformation for the propagation path} [O,\overline{O}] $ \\
            $12: \textbf{Return } \mathcal{G}$ \\

		\bottomrule	
	\end{tabular}
	\label{alg:obf_coupling}
	\vspace{-0.5em}
\end{table}

\begin{table*}[t]
\small
\centering
\caption{The scaled maximal element-wise error of our proposed \ObfuscationToolname compared with the existing model obfuscation method ModelObfuscator~\cite{zhou2023modelobfuscator}.}
\begin{tabular}{lcccccccccc|c}
\hline
 Model name    & Fruit & Skin cancer &MobileNet&MNASNet &SqueezeNet &EfficientNet &MiDaS &Lenet &PoseNet &SSD  & Average \\
\cline{1-12}
 ModelObfuscator     & 0.0 & 0.0 & 0.0      & 0.0   &  0.0    & 0.0    & 0.0  & 0.0        &  0.0   & 0.0    & ${\bf 0.0}$ \\
 \textbf{\ObfuscationToolname}  & $1.4\!\times\!10^{-7}$  & $2.0\!\times\!10^{-7}$ & $4.9\!\times\!10^{-9}$    & $5.9\!\times\!10^{-8}$   & $3.8\!\times\!10^{-9}$      & $1.2\!\times\!10^{-8}$    &  $3.8\!\times\!10^{-7}$ &  $4.8\!\times\!10^{-7}$       &$4.3\!\times\!10^{-8}$  &  $6.5\!\times\!10^{-8}$   & $ {\bf 1.4\!\times\!10^{-7}}$\\
\hline
\end{tabular}
\label{tb:obf_acc}
\end{table*}

\begin{table*}[t]
\small
\centering
\caption{Performance of \ObfuscationToolname in defending against the instrumentation attack \InstrumentationToolname. `TN': True negative (\ie correct identification for obfuscating operators) rate of operator classification. `Difference': the average value difference between the attacking performance based on \ObfuscationToolname (this table) and existing obfuscation methods (Table~\ref{tb:deobf}). `WER, NIR, OCA, SS': the lower is better. `WEE': the higher is better. }
\begin{tabular}{lcccccccccc|c|c}
\hline
 Metric   & Fruit   & Skin     &MobileNet    &MNASNet   &SqueezeNet &EfficientNet &MiDaS     &LeNet     &PoseNet  &SSD        & \textbf{Average value} & \textbf{Difference} \\
\hline
WER      & 55.17\%  & 51.72\%  & 35.71\%     & 50.00\%  & 62.96\%   & 68.63\%     &  56.94\% &25.00\%   & 59.38\% & 59.72\%   & \textbf{52.52\%}   & \textbf{46.24} $\downarrow$ \\
NIR      & 52.54\%  & 51.67\%  & 49.18\%     & 70.45\%  & 74.70\%   & 73.81\%     &  71.26\%   & 33.33\%  & 26.19\% &  72.82\%  & \textbf{57.60\%}   & \textbf{40.80\%} $\downarrow$    \\
OCA      & 52.54\%  & 51.67\%  & 50.82\%     & 70.79\%  & 73.26\%   & 73.81\%     &  79.31\% &  29.63\% & 57.89\% &   80.65\% & \textbf{60.04\%}   & \textbf{37.85\%} $\downarrow$    \\
TN (OCA) &    0\%   & 0\%      & 0\%         & 0\%      & 0\%       & 0\%     &  0\%     &  5.00\%  & 0\%     &   0\%     & \textbf{0.5\%}    & N/A   \\
\hline
WEE      & 0.61     & 1.46     & 2.70        & 1.25     & 0.32      & 0.33        &  0.14    & 0.04     & 0.75    &  0.24     & \textbf{0.78}      &  \textbf{0.78}    $\uparrow$     \\
SS      &  0.53     & 0.52     & 0.49        & 0.70     & 0.75      & 0.74        &  0.71    & 0.33    & 0.26    &  0.73     & \textbf{0.58}      &  \textbf{0.40}    $\downarrow$     \\
\hline
\end{tabular}
\label{tb:deobf_mod}
\end{table*}

\paragraph{\textbf{Algorithm}} Our \ObfuscationToolname will first use existing obfuscation methods~\cite{zhou2023modelobfuscator} to obfuscate the model representation. Then, it will perform the Obfuscation Coupling algorithm, which is shown in Algorithm~1. Note that our dynamic obfuscation strategy uses linear transformation, it supports obfuscating the same operators several times. So, \ObfuscationToolname will randomly sample $n$ obfuscation pairs (we set the $n$ to the total number of operators) from $\mathcal{G}$ based on the Lemma~\ref{lemma_1} and Theorem~\ref{theorem_1}. Note that the selected operator and the coupled operator may include multiple operators. For example, as shown in Figure~\ref{fig:mod}, if the green \texttt{conv2d} is chosen as the selected operator, the coupled operator can be \texttt{mjzdmh} and \texttt{fsebwa} as they are both the next operator of the \texttt{conv2d}. Finally, \ObfuscationToolname will follow the rule defined in Lemma~\ref{lemma_1} and Theorem~\ref{theorem_1} to obfuscate the weights through a linear transformation. Such transformation is hard to detect unless attackers know the original weights.

\subsection{Compilation}
After obtaining the obfuscated model and the modified source code of the API library using an existing DL obfuscation tool~\cite{zhou2023modelobfuscator}, \ObfuscationToolname then assembles the new obfuscated model generated by its dynamic obfuscation strategy in Python. Then, it recompiles the modified TFLite library to support the newly generated obfuscated model. The newly generated obfuscated model and library can be packaged into mobile apps or embedded device software to replace the original unobfuscated model or obfuscated model generated by existing obfuscation strategies.

\subsection{\ObfuscationToolname Effectiveness}

Note that in our evaluations, we set the number of obfuscation pairs $n$ in Algorithm~1 to the total number of operators. The settings are the same as the evaluation in section~\ref{sec:deobf}. \textbf{Note that we set the number of extra obfuscating operators to 30 in evaluating the effectiveness of \ObfuscationToolname.}

\subsubsection{\textbf{Output Difference}}
We need to first evaluate the performance loss of our proposed obfuscation strategy. Table~\ref{tb:obf_acc} summarises the evaluation of our methods on performance loss compared with existing model obfuscation~\cite{zhou2023modelobfuscator}. The maximal element-wise error can be formulated as:
\begin{equation}
    \theta = (\max _{i=1}^N|\mathcal{C}\left(x_i\right)-\mathcal{G}\left(x_i\right)|) \div \max _{i=1}^N|\mathcal{G}\left(x_i\right)|
\end{equation}
where $\mathcal{G}$ is the model results before obfuscation. $\mathcal{C}$ refers to the output of obfuscated models. $N$ denote the total number of output elements (the output is usually a matrix or vector). We use 100 inputs to compute the maximal element-wise error and divide the error by the maximal value of the original output to get the scaled maximal element-wise error.
Our \ObfuscationToolname method only has negligible errors, because \ObfuscationToolname theoretically uses the same computing process as the original model inference. But the weights transformation process in our \ObfuscationToolname will have an inevitable computing error that is very small.

\begin{table}[t]
 \footnotesize
\centering
\caption{The ability to resist the reverse engineering for on-device models. `$\surd$': this model parsing method cannot extract information for all models. `Model conversion': TF-ONNX~\citep{tf2onnx}, TFLite2ONNX~\citep{tflite2onnx}, and TFLite2TF~\citep{tflite2tensorflow}. `Parsing in buffer': ~\cite{li2021deeppayload}. `Feature analyzing': ~\cite{huang2022smart}.} 
\begin{tabular}{l|c|c|c}
\toprule
                                       &  {\bf Model conversion} & {\bf Parsing in buffer}  & {\bf Feature analyzing} \\
\hline
ModelObfuscator     & $\surd$                       & $\surd$            & $\surd$          \\ 
\textbf{\ObfuscationToolname}     & $\surd$                   & $\surd$            & $\surd$          \\ 
\hline
\end{tabular}
\label{tb:extraction}
\end{table}

\begin{table*}[t!]
\centering
\caption{Time overhead (seconds per 1000 inputs) of \ObfuscationToolname compared with existing obfuscation method~\cite{zhou2023modelobfuscator} on \texttt{x86-64} and \texttt{ARM64} platforms. We set the number of extra layers to 20 for both obfuscation strategies. }
\begin{tabular}{lcccccccccc|c}
\hline
& \multicolumn{10}{c}{\texttt{x86-64}} & \\
\hline
                     & Fruit & Skin &MobileNet&MNASNet &SqueezeNet &EffcientNet &MiDaS   &LeNet    &PoseNet &SSD     & \textbf{Average value} \\
\hline
original             & 28.6  & 86.7 & 53.3    & 69.4   & 37.5      & 99.7       &  309.1 &  2.1    & 114.3  & 208.6  & \textbf{100.9}        \\
\hline
ModelObfuscator      & 28.4  & 87.4 & 55.6    & 71.2   & 39.7      & 101.1      &  321.7 &  2.9    & 117.8  & 211.1  & \textbf{103.7}       \\

\ObfuscationToolname & 29.1  & 87.7 & 54.9    & 71.5   & 40.3      & 102.4      &  329.2 &  2.9    & 117.9  & 212.6      & \textbf{104.9}       \\
\hline
& \multicolumn{10}{c}{\texttt{ARM64}} & \\
\hline
                     & Fruit & Skin &MobileNet&MNASNet &SqueezeNet &EffcientNet &MiDaS   &LeNet    &PoseNet &SSD     & \textbf{Average value} \\
\hline
original             & 14.2 & 45.5  & 28.1     &  36.0  &  53.2   &  41.1   & 380.9  & 4.6  &  43.8   & 92.3  & \textbf{73.9}        \\
\hline
ModelObfuscator      & 14.3  & 45.9 & 28.6    & 36.2   & 53.7      & 41.1      &  381.4 &  4.9    & 44.1  & 92.4  & \textbf{74.3}       \\

\ObfuscationToolname & 14.2  & 46.2 & 28.8    & 36.5   & 53.8      & 41.4      &  385.3 &  4.9    & 44.7  & 92.6      & \textbf{74.8}       \\
\hline

\end{tabular}
\label{tb:latency}
\end{table*}

\begin{table*}[t!]
\centering
\caption{Overhead of \ObfuscationToolname on random access memory (RAM) cost (Mb per model) compared with existing obfuscation method~\cite{zhou2023modelobfuscator}. To eliminate the influence of other processes on the test machine, we show the increment of RAM usage.}
\begin{tabular}{lcccccccccc|c}
\hline
                     & Fruit  & Skin    &MobileNet &MNASNet &SqueezeNet &EffcientNet  & MiDaS    &LeNet   &PoseNet  &  SSD    & \textbf{Average value} \\
\hline
Original             & 13.5   & 34.6    & 23.9     & 37.7   & 34.4      & 43.4        &  253.8   & 6.2    & 38.8    &  97.8   &  \textbf{58.4}       \\
\hline
ModelObfuscator      & 22.2   & 54.9    & 38.0     & 56.6   & 47.9      & 59.6        &  278.9   & 6.8    & 55.4    &  107.6  &  \textbf{72.8}       \\

\ObfuscationToolname & 26.9   &  58.5   & 39.2     & 56.3   & 48.6      & 61.1        &  281.4   & 9.3    & 58.5    & 108.3   &  \textbf{74.8}       \\

\hline
\end{tabular}
\label{tb:mem}
\end{table*}

\subsubsection{\textbf{Resilience to Attack}}
The effectiveness evaluation of our \ObfuscationToolname is shown in Table~\ref{tb:deobf}. Compared with the existing model obfuscation method (see Table~\ref{tb:deobf}), the \InstrumentationToolname significantly reduce the performance of attacks based on dynamic instrumentation. Note that we only introduce a maximum of 30 obfuscating operators to the obfuscated model. Because \InstrumentationToolname will classify all operators (including valid operators and obfuscating operators) to valid operators, \InstrumentationToolname achieves $60.04\%$ of Operator Classification Accuracy (OCA). It only obtains $0.5\%$ of true negative rate that shows the performance of correctly identifying the obfuscating operator. Our method can obfuscate the intermediate results, the \InstrumentationToolname only achieves $57.60\%$ of the operator's name identification rate (NIR). Our \ObfuscationToolname also achieves high performance on weights obfuscation, attackers can only correctly extract $52.52\%$ of model weights. In global metrics (\ie WEE and SS), the extracted weights have high errors (\ie 0.78) compared with the results of existing obfuscation methods (\ie $2.1 \times 10^{-5}$). Our method also can significantly decrease the structure similarity (SS) from 0.98 to 0.58.
The results show the dynamic instrumentation analysis method cannot effectively deobfuscate the models generated by our proposed dynamic obfuscation strategy. Note that The model representations are randomly obfuscated by our method, it is hard for attackers to identify which operators have been obfuscated. Therefore, we do not need to apply our \ObfuscationToolname to obfuscate all model weights, operators, and intermediate results to reduce the performance loss.

In addition, we follow the setting in~\cite{zhou2023modelobfuscator} to check whether our proposed obfuscation strategy can still be robust for the reverse engineering of an on-device model or not. The results are shown in~\ref{tb:extraction}. It shows our proposed strategy will not degrade the model resistance against normal reverse engineering methods.

\begin{tcolorbox}[colback=gray!5!white,colframe=gray!85]
RQ2: Our \ObfuscationToolname is effective in defending against existing model parsing methods and dynamic instrumentation attack method \InstrumentationToolname, with only negligible performance loss.
\end{tcolorbox}

\section{RQ3: Efficiency of \ObfuscationToolname}

As \ObfuscationToolname introduces additional computations via its extra obfuscating operators, it is important to evaluate the influence of such additional computation on inference efficiency. In order to evaluate the efficiency impact of our obfuscation strategies on ML models, we conducted experiments to measure new obfuscated ML model runtime overhead. 

%

\paragraph{\textbf{Experimental Environment:}}
The efficiency of \ObfuscationToolname is evaluated on a workstation with Intel(R) Xeon(R) W-2175 2.50GHz CPU, 32GB RAM, with Ubuntu 20.04.1 operating system and a Xiaomi 11 Pro smartphone with Android 13 OS.


\paragraph{\textbf{Time Overhead:}} We measured the time overhead of \ObfuscationToolname and existing DL model obfuscation strategies \cite{zhou2023modelobfuscator}, based on 5,000 randomly generated instances. The results of these experiments are presented in Tables \ref{tb:latency}, which report the time overhead of the obfuscated models.  As shown in Table~\ref{tb:latency}, even though additional computations are introduced into the extra obfuscating operators, \textbf{\ObfuscationToolname obfuscated models incur a negligible time overhead compared with existing model obfuscation strategy}. This is because the obfuscating operators just perform simple linear transformations, which have low complexity compared with the whole model inference process.

\paragraph{\textbf{RAM Overhead:}}
We measured the RAM overhead of both obfuscation strategies. The memory overhead for \ObfuscationToolname obfuscated models is shown in Table~\ref{tb:mem}. To eliminate the impact of different memory optimization methods, like the study~\cite{zhou2023modelobfuscator}, we use peak RAM usage where the model preserves all intermediate tensors. 
\textbf{Our method only introduces negligible overheads compared with the existing model obfuscation strategies}. Because our obfuscation strategy only transforms the value of the weights.

\begin{tcolorbox}[colback=gray!5!white,colframe=gray!85]
RQ3: The time and RAM overhead of \ObfuscationToolname are both negligible, while RQ2 shows that our proposed \ObfuscationToolname strategy significantly improves the security level of the ob-device models.
\end{tcolorbox}

\section{Limitations}

Our proposed \InstrumentationToolname and \ObfuscationToolname are designed for on-device TFLite models. Although our strategy can be adapted to other DL platforms, we do not know if there are unsupported DL model constructs our approach may not be able to support.


We have not evaluated our \ObfuscationToolname with real-world mobile ML developers. 
We have not evaluated our \ObfuscationToolname with side-channel information (\eg RAM, CPU usage) to reconstruct the model~\cite{wei2020leaky,duddu2018stealing,liu2019side}. Our proposed \ObfuscationToolname may not be effective for them because the obfuscated model has a similar resource consumption pattern to the original models. However, our method is designed for and very effective for resisting the attacks based-on program analysis. 

Our dynamic obfuscation method \ObfuscationToolname will cause a very slight performance loss for on-device models. In addition, our proposed \ObfuscationToolname will introduce very small overheads to the model inference similar to the existing model obfuscation strategies.

\paragraph{\textbf{Threats to Validity}} For internal threats to validity, the efficiency evaluation results (RQ3) may be affected by other services running in the experimental environments (i.e., Ubuntu server, Xiaomi 11 Pro). For external threats to validity, as the DL techniques and AI compilers may change a lot, our tool needs continuous updates in the future.

\begin{figure}[h!]
  \begin{center}
    \includegraphics[width=0.9\linewidth]{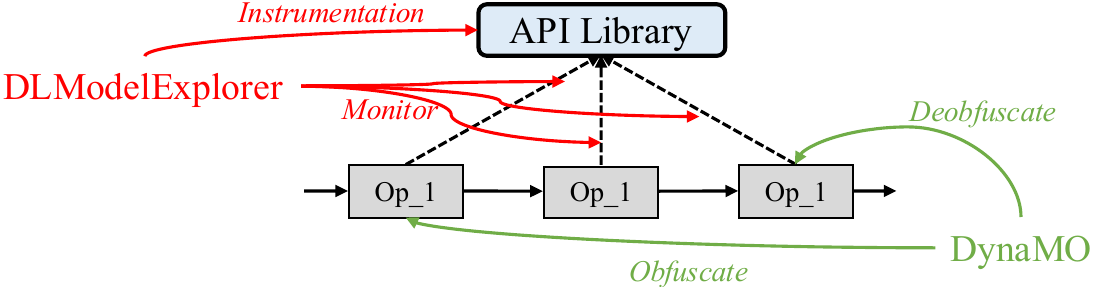}
  \end{center}
  \caption{Meta-model our method.}
  \label{fig:meta}
\end{figure}

\paragraph{\textbf{Meta-Model}} Although we focus on the TFLite because it's the most commonly used DL platform on mobile devices, especially for Android, our methods are also general for other DL compilers. We propose two methods: \InstrumentationToolname and \ObfuscationToolname. \InstrumentationToolname will monitor the APIs (DL compilers are usually open-sourced) that allocate the model weights to the memory and obtain the intermediate results by instrumentation to predict the operator's type. \ObfuscationToolname only uses linear weight transformations to obfuscate and restore the model's outputs. The meta-model of our study is shown in Figure~\ref{fig:meta}.




\section{Conclusion}

We analyzed the risk of deep learning models deployed on mobile devices and the vulnerabilities of the existing static and half-dynamic DL model obfuscation strategies. We showed that attackers can still extract information from these obfuscated models using dynamic instrumentation-based techniques. To address this vulnerability, we propose a novel dynamic model obfuscation strategy and tool, \ObfuscationToolname to better secure mobile device deployed DL models. \ObfuscationToolname can couple the obfuscated DL model operators to increase the randomness of the obfuscation process without significant overhead. In addition, we provide a theoretical guarantee to achieve such dynamic obfuscation without model performance loss. We developed a prototype tool \ObfuscationToolname to automatically obfuscate TFLite DL models using our proposed obfuscation strategy.
Experiments show that our method is effective in resisting the model parsing tools and the proposed dynamic instrumentation attack without performance sacrifice. In the future, we will optimize the obfuscation process and our prototype tool to further increase the efficiency of our method and tool.

\section*{Acknowledgements}

This work is partially supported by the Open Foundation of Yunnan Key Laboratory of Software Engineering under Grant No.2023SE102. Zhou is supported by a Faculty of IT PhD scholarship. Grundy is supported by ARC Laureate Fellowship FL190100035.

\bibliographystyle{ACM-Reference-Format}
\bibliography{acmart}


\begin{thebibliography}{39}


\ifx \showCODEN    \undefined \def \showCODEN     #1{\unskip}     \fi
\ifx \showDOI      \undefined \def \showDOI       #1{#1}\fi
\ifx \showISBNx    \undefined \def \showISBNx     #1{\unskip}     \fi
\ifx \showISBNxiii \undefined \def \showISBNxiii  #1{\unskip}     \fi
\ifx \showISSN     \undefined \def \showISSN      #1{\unskip}     \fi
\ifx \showLCCN     \undefined \def \showLCCN      #1{\unskip}     \fi
\ifx \shownote     \undefined \def \shownote      #1{#1}          \fi
\ifx \showarticletitle \undefined \def \showarticletitle #1{#1}   \fi
\ifx \showURL      \undefined \def \showURL       {\relax}        \fi
\providecommand\bibfield[2]{#2}
\providecommand\bibinfo[2]{#2}
\providecommand\natexlab[1]{#1}
\providecommand\showeprint[2][]{arXiv:#2}

\bibitem[tf2(2022)]%
        {tf2onnx}
 \bibinfo{year}{2022}\natexlab{}.
\newblock \bibinfo{booktitle}{\emph{{tf2onnx - Convert TensorFlow, Keras,
  Tensorflow.js and Tflite models to ONN}}}.
\newblock
\urldef\tempurl%
\url{https://github.com/onnx/tensorflow-onnx}
\showURL{%
\tempurl}


\bibitem[apk(2024)]%
        {apktool}
 \bibinfo{year}{2024}\natexlab{}.
\newblock \bibinfo{booktitle}{\emph{{Apktool: A tool for reverse engineering
  Android apk files}}}.
\newblock
\urldef\tempurl%
\url{https://ibotpeaches.github.io/Apktool/}
\showURL{%
\tempurl}


\bibitem[min(2024)]%
        {mindspore}
 \bibinfo{year}{2024}\natexlab{}.
\newblock \bibinfo{booktitle}{\emph{{Mindspore}}}.
\newblock
\urldef\tempurl%
\url{https://www.mindspore.cn/lite/docs/en/r1.7/use/obfuscator_tool.html}
\showURL{%
\tempurl}


\bibitem[pin(2024)]%
        {pin}
 \bibinfo{year}{2024}\natexlab{}.
\newblock \bibinfo{booktitle}{\emph{{Pin: A Dynamic Binary Instrumentation
  Tool}}}.
\newblock
\urldef\tempurl%
\url{https://www.intel.com/content/www/us/en/developer/articles/tool/pin-a-dynamic-binary-instrumentation-tool.html}
\showURL{%
\tempurl}


\bibitem[Abadi et~al\mbox{.}(2015)]%
        {tensorflow2015_whitepaper}
\bibfield{author}{\bibinfo{person}{Mart\'{i}n Abadi}, \bibinfo{person}{Ashish
  Agarwal}, \bibinfo{person}{Paul Barham}, \bibinfo{person}{Eugene Brevdo},
  \bibinfo{person}{Zhifeng Chen}, \bibinfo{person}{Craig Citro},
  \bibinfo{person}{Greg~S. Corrado}, \bibinfo{person}{Andy Davis},
  \bibinfo{person}{Jeffrey Dean}, \bibinfo{person}{Matthieu Devin},
  \bibinfo{person}{Sanjay Ghemawat}, \bibinfo{person}{Ian Goodfellow},
  \bibinfo{person}{Andrew Harp}, \bibinfo{person}{Geoffrey Irving},
  \bibinfo{person}{Michael Isard}, \bibinfo{person}{Yangqing Jia},
  \bibinfo{person}{Rafal Jozefowicz}, \bibinfo{person}{Lukasz Kaiser},
  \bibinfo{person}{Manjunath Kudlur}, \bibinfo{person}{Josh Levenberg},
  \bibinfo{person}{Dandelion Man\'{e}}, \bibinfo{person}{Rajat Monga},
  \bibinfo{person}{Sherry Moore}, \bibinfo{person}{Derek Murray},
  \bibinfo{person}{Chris Olah}, \bibinfo{person}{Mike Schuster},
  \bibinfo{person}{Jonathon Shlens}, \bibinfo{person}{Benoit Steiner},
  \bibinfo{person}{Ilya Sutskever}, \bibinfo{person}{Kunal Talwar},
  \bibinfo{person}{Paul Tucker}, \bibinfo{person}{Vincent Vanhoucke},
  \bibinfo{person}{Vijay Vasudevan}, \bibinfo{person}{Fernanda Vi\'{e}gas},
  \bibinfo{person}{Oriol Vinyals}, \bibinfo{person}{Pete Warden},
  \bibinfo{person}{Martin Wattenberg}, \bibinfo{person}{Martin Wicke},
  \bibinfo{person}{Yuan Yu}, {and} \bibinfo{person}{Xiaoqiang Zheng}.}
  \bibinfo{year}{2015}\natexlab{}.
\newblock \bibinfo{title}{{TensorFlow}: Large-Scale Machine Learning on
  Heterogeneous Systems}.
\newblock
\newblock
\urldef\tempurl%
\url{https://www.tensorflow.org/}
\showURL{%
\tempurl}
\newblock
\shownote{Software available from tensorflow.org}.


\bibitem[Chen et~al\mbox{.}(2022)]%
        {chen2022learning}
\bibfield{author}{\bibinfo{person}{Simin Chen}, \bibinfo{person}{Hamed
  Khanpour}, \bibinfo{person}{Cong Liu}, {and} \bibinfo{person}{Wei Yang}.}
  \bibinfo{year}{2022}\natexlab{}.
\newblock \showarticletitle{Learning to reverse dnns from ai programs
  automatically}. In \bibinfo{booktitle}{\emph{AAAI Conference on Artificial
  Intelligence}}.
\newblock


\bibitem[Chollet et~al\mbox{.}(2018)]%
        {chollet2018keras}
\bibfield{author}{\bibinfo{person}{Fran{\c{c}}ois Chollet} {et~al\mbox{.}}}
  \bibinfo{year}{2018}\natexlab{}.
\newblock \showarticletitle{Keras: The python deep learning library}.
\newblock \bibinfo{journal}{\emph{Astrophysics source code library}}
  (\bibinfo{year}{2018}), \bibinfo{pages}{ascl--1806}.
\newblock


\bibitem[Collberg et~al\mbox{.}(1997)]%
        {collberg1997taxonomy}
\bibfield{author}{\bibinfo{person}{Christian Collberg}, \bibinfo{person}{Clark
  Thomborson}, {and} \bibinfo{person}{Douglas Low}.}
  \bibinfo{year}{1997}\natexlab{}.
\newblock \bibinfo{title}{A taxonomy of obfuscating transformations}.
\newblock
\newblock


\bibitem[Collberg et~al\mbox{.}(1998)]%
        {collberg1998manufacturing}
\bibfield{author}{\bibinfo{person}{Christian Collberg}, \bibinfo{person}{Clark
  Thomborson}, {and} \bibinfo{person}{Douglas Low}.}
  \bibinfo{year}{1998}\natexlab{}.
\newblock \showarticletitle{Manufacturing cheap, resilient, and stealthy opaque
  constructs}. In \bibinfo{booktitle}{\emph{Proceedings of the 25th ACM
  SIGPLAN-SIGACT symposium on Principles of programming languages}}.
  \bibinfo{pages}{184--196}.
\newblock
\urldef\tempurl%
\url{https://doi.org/10.1145/268946.268962}
\showDOI{\tempurl}


\bibitem[Dilhara et~al\mbox{.}(2021)]%
        {dilhara2021understanding}
\bibfield{author}{\bibinfo{person}{Malinda Dilhara}, \bibinfo{person}{Ameya
  Ketkar}, {and} \bibinfo{person}{Danny Dig}.} \bibinfo{year}{2021}\natexlab{}.
\newblock \showarticletitle{Understanding Software-2.0: A Study of Machine
  Learning library usage and evolution}.
\newblock \bibinfo{journal}{\emph{ACM Transactions on Software Engineering and
  Methodology (TOSEM)}} \bibinfo{volume}{30}, \bibinfo{number}{4}
  (\bibinfo{year}{2021}), \bibinfo{pages}{1--42}.
\newblock
\urldef\tempurl%
\url{https://doi.org/10.1145/3453478}
\showDOI{\tempurl}


\bibitem[Duddu et~al\mbox{.}(2018)]%
        {duddu2018stealing}
\bibfield{author}{\bibinfo{person}{Vasisht Duddu}, \bibinfo{person}{Debasis
  Samanta}, \bibinfo{person}{D~Vijay Rao}, {and} \bibinfo{person}{Valentina~E
  Balas}.} \bibinfo{year}{2018}\natexlab{}.
\newblock \showarticletitle{Stealing neural networks via timing side channels}.
\newblock \bibinfo{journal}{\emph{arXiv preprint arXiv:1812.11720}}
  (\bibinfo{year}{2018}).
\newblock


\bibitem[Howard et~al\mbox{.}(2017)]%
        {howard2017mobilenets}
\bibfield{author}{\bibinfo{person}{Andrew~G Howard}, \bibinfo{person}{Menglong
  Zhu}, \bibinfo{person}{Bo Chen}, \bibinfo{person}{Dmitry Kalenichenko},
  \bibinfo{person}{Weijun Wang}, \bibinfo{person}{Tobias Weyand},
  \bibinfo{person}{Marco Andreetto}, {and} \bibinfo{person}{Hartwig Adam}.}
  \bibinfo{year}{2017}\natexlab{}.
\newblock \showarticletitle{Mobilenets: Efficient convolutional neural networks
  for mobile vision applications}.
\newblock \bibinfo{journal}{\emph{arXiv preprint arXiv:1704.04861}}
  (\bibinfo{year}{2017}).
\newblock


\bibitem[Huang and Chen(2022)]%
        {huang2022smart}
\bibfield{author}{\bibinfo{person}{Yujin Huang} {and} \bibinfo{person}{Chunyang
  Chen}.} \bibinfo{year}{2022}\natexlab{}.
\newblock \showarticletitle{Smart App Attack: Hacking Deep Learning Models in
  Android Apps}.
\newblock \bibinfo{journal}{\emph{IEEE Transactions on Information Forensics
  and Security}}  \bibinfo{volume}{17} (\bibinfo{year}{2022}),
  \bibinfo{pages}{1827--1840}.
\newblock


\bibitem[Hyodo(2022)]%
        {tflite2tensorflow}
\bibfield{author}{\bibinfo{person}{Katsuya Hyodo}.}
  \bibinfo{year}{2022}\natexlab{}.
\newblock \bibinfo{booktitle}{\emph{{tflite2tensorflow}}}.
\newblock
\urldef\tempurl%
\url{https://github.com/PINTO0309/tflite2tensorflow}
\showURL{%
\tempurl}


\bibitem[Iandola et~al\mbox{.}(2016)]%
        {iandola2016squeezenet}
\bibfield{author}{\bibinfo{person}{Forrest~N Iandola}, \bibinfo{person}{Song
  Han}, \bibinfo{person}{Matthew~W Moskewicz}, \bibinfo{person}{Khalid Ashraf},
  \bibinfo{person}{William~J Dally}, {and} \bibinfo{person}{Kurt Keutzer}.}
  \bibinfo{year}{2016}\natexlab{}.
\newblock \showarticletitle{SqueezeNet: AlexNet-level accuracy with 50x fewer
  parameters and< 0.5 MB model size}.
\newblock \bibinfo{journal}{\emph{arXiv preprint arXiv:1602.07360}}
  (\bibinfo{year}{2016}).
\newblock


\bibitem[Kendall et~al\mbox{.}(2015)]%
        {kendall2015posenet}
\bibfield{author}{\bibinfo{person}{Alex Kendall}, \bibinfo{person}{Matthew
  Grimes}, {and} \bibinfo{person}{Roberto Cipolla}.}
  \bibinfo{year}{2015}\natexlab{}.
\newblock \showarticletitle{Posenet: A convolutional network for real-time
  6-dof camera relocalization}. In \bibinfo{booktitle}{\emph{Proceedings of the
  IEEE international conference on computer vision}}.
  \bibinfo{pages}{2938--2946}.
\newblock
\urldef\tempurl%
\url{https://doi.org/10.1109/iccv.2015.336}
\showDOI{\tempurl}


\bibitem[LeCun et~al\mbox{.}(1998)]%
        {lecun1998gradient}
\bibfield{author}{\bibinfo{person}{Yann LeCun}, \bibinfo{person}{L{\'e}on
  Bottou}, \bibinfo{person}{Yoshua Bengio}, {and} \bibinfo{person}{Patrick
  Haffner}.} \bibinfo{year}{1998}\natexlab{}.
\newblock \showarticletitle{Gradient-based learning applied to document
  recognition}.
\newblock \bibinfo{journal}{\emph{Proc. IEEE}} \bibinfo{volume}{86},
  \bibinfo{number}{11} (\bibinfo{year}{1998}), \bibinfo{pages}{2278--2324}.
\newblock
\urldef\tempurl%
\url{https://doi.org/10.1109/5.726791}
\showDOI{\tempurl}


\bibitem[Li et~al\mbox{.}(2021)]%
        {li2021deeppayload}
\bibfield{author}{\bibinfo{person}{Yuanchun Li}, \bibinfo{person}{Jiayi Hua},
  \bibinfo{person}{Haoyu Wang}, \bibinfo{person}{Chunyang Chen}, {and}
  \bibinfo{person}{Yunxin Liu}.} \bibinfo{year}{2021}\natexlab{}.
\newblock \showarticletitle{Deeppayload: Black-box backdoor attack on deep
  learning models through neural payload injection}. In
  \bibinfo{booktitle}{\emph{2021 IEEE/ACM 43rd International Conference on
  Software Engineering (ICSE)}}. IEEE, \bibinfo{pages}{263--274}.
\newblock
\urldef\tempurl%
\url{https://doi.org/10.1109/icse43902.2021.00035}
\showDOI{\tempurl}


\bibitem[Liu et~al\mbox{.}(2019)]%
        {liu2019side}
\bibfield{author}{\bibinfo{person}{Sihang Liu}, \bibinfo{person}{Yizhou Wei},
  \bibinfo{person}{Jianfeng Chi}, \bibinfo{person}{Faysal~Hossain Shezan},
  {and} \bibinfo{person}{Yuan Tian}.} \bibinfo{year}{2019}\natexlab{}.
\newblock \showarticletitle{Side channel attacks in computation offloading
  systems with gpu virtualization}. In \bibinfo{booktitle}{\emph{2019 IEEE
  Security and Privacy Workshops (SPW)}}. IEEE, \bibinfo{pages}{156--161}.
\newblock


\bibitem[Liu et~al\mbox{.}(2016)]%
        {liu2016ssd}
\bibfield{author}{\bibinfo{person}{Wei Liu}, \bibinfo{person}{Dragomir
  Anguelov}, \bibinfo{person}{Dumitru Erhan}, \bibinfo{person}{Christian
  Szegedy}, \bibinfo{person}{Scott Reed}, \bibinfo{person}{Cheng-Yang Fu},
  {and} \bibinfo{person}{Alexander~C Berg}.} \bibinfo{year}{2016}\natexlab{}.
\newblock \showarticletitle{Ssd: Single shot multibox detector}. In
  \bibinfo{booktitle}{\emph{European conference on computer vision}}. Springer,
  \bibinfo{pages}{21--37}.
\newblock


\bibitem[Papernot et~al\mbox{.}(2017)]%
        {papernot2017practical}
\bibfield{author}{\bibinfo{person}{Nicolas Papernot}, \bibinfo{person}{Patrick
  McDaniel}, \bibinfo{person}{Ian Goodfellow}, \bibinfo{person}{Somesh Jha},
  \bibinfo{person}{Z~Berkay Celik}, {and} \bibinfo{person}{Ananthram Swami}.}
  \bibinfo{year}{2017}\natexlab{}.
\newblock \showarticletitle{Practical black-box attacks against machine
  learning}. In \bibinfo{booktitle}{\emph{Proceedings of the 2017 ACM on Asia
  conference on computer and communications security}}.
  \bibinfo{pages}{506--519}.
\newblock


\bibitem[Paszke et~al\mbox{.}(2019)]%
        {paszke2019pytorch}
\bibfield{author}{\bibinfo{person}{Adam Paszke}, \bibinfo{person}{Sam Gross},
  \bibinfo{person}{Francisco Massa}, \bibinfo{person}{Adam Lerer},
  \bibinfo{person}{James Bradbury}, \bibinfo{person}{Gregory Chanan},
  \bibinfo{person}{Trevor Killeen}, \bibinfo{person}{Zeming Lin},
  \bibinfo{person}{Natalia Gimelshein}, \bibinfo{person}{Luca Antiga},
  {et~al\mbox{.}}} \bibinfo{year}{2019}\natexlab{}.
\newblock \showarticletitle{Pytorch: An imperative style, high-performance deep
  learning library}.
\newblock \bibinfo{journal}{\emph{Advances in neural information processing
  systems}}  \bibinfo{volume}{32} (\bibinfo{year}{2019}).
\newblock


\bibitem[Ranftl et~al\mbox{.}(2020)]%
        {ranftl2020towards}
\bibfield{author}{\bibinfo{person}{Ren{\'e} Ranftl}, \bibinfo{person}{Katrin
  Lasinger}, \bibinfo{person}{David Hafner}, \bibinfo{person}{Konrad
  Schindler}, {and} \bibinfo{person}{Vladlen Koltun}.}
  \bibinfo{year}{2020}\natexlab{}.
\newblock \showarticletitle{Towards robust monocular depth estimation: Mixing
  datasets for zero-shot cross-dataset transfer}.
\newblock \bibinfo{journal}{\emph{IEEE transactions on pattern analysis and
  machine intelligence}} \bibinfo{volume}{44}, \bibinfo{number}{3}
  (\bibinfo{year}{2020}), \bibinfo{pages}{1623--1637}.
\newblock
\urldef\tempurl%
\url{https://doi.org/10.1109/tpami.2020.3019967}
\showDOI{\tempurl}


\bibitem[Ren et~al\mbox{.}(2024)]%
        {ren2024demistify}
\bibfield{author}{\bibinfo{person}{Pengcheng Ren}, \bibinfo{person}{Chaoshun
  Zuo}, \bibinfo{person}{Xiaofeng Liu}, \bibinfo{person}{Wenrui Diao},
  \bibinfo{person}{Qingchuan Zhao}, {and} \bibinfo{person}{Shanqing Guo}.}
  \bibinfo{year}{2024}\natexlab{}.
\newblock \showarticletitle{DEMISTIFY: Identifying On-device Machine Learning
  Models Stealing and Reuse Vulnerabilities in Mobile Apps}. In
  \bibinfo{booktitle}{\emph{Proceedings of the 46th IEEE/ACM International
  Conference on Software Engineering}}. \bibinfo{pages}{1--13}.
\newblock


\bibitem[Schrittwieser et~al\mbox{.}(2016)]%
        {schrittwieser2016protecting}
\bibfield{author}{\bibinfo{person}{Sebastian Schrittwieser},
  \bibinfo{person}{Stefan Katzenbeisser}, \bibinfo{person}{Johannes Kinder},
  \bibinfo{person}{Georg Merzdovnik}, {and} \bibinfo{person}{Edgar Weippl}.}
  \bibinfo{year}{2016}\natexlab{}.
\newblock \showarticletitle{Protecting software through obfuscation: Can it
  keep pace with progress in code analysis?}
\newblock \bibinfo{journal}{\emph{ACM Computing Surveys (CSUR)}}
  \bibinfo{volume}{49}, \bibinfo{number}{1} (\bibinfo{year}{2016}),
  \bibinfo{pages}{1--37}.
\newblock
\urldef\tempurl%
\url{https://doi.org/10.1145/2886012}
\showDOI{\tempurl}


\bibitem[Tan et~al\mbox{.}(2019)]%
        {tan2019mnasnet}
\bibfield{author}{\bibinfo{person}{Mingxing Tan}, \bibinfo{person}{Bo Chen},
  \bibinfo{person}{Ruoming Pang}, \bibinfo{person}{Vijay Vasudevan},
  \bibinfo{person}{Mark Sandler}, \bibinfo{person}{Andrew Howard}, {and}
  \bibinfo{person}{Quoc~V Le}.} \bibinfo{year}{2019}\natexlab{}.
\newblock \showarticletitle{Mnasnet: Platform-aware neural architecture search
  for mobile}. In \bibinfo{booktitle}{\emph{Proceedings of the IEEE/CVF
  Conference on Computer Vision and Pattern Recognition}}.
  \bibinfo{pages}{2820--2828}.
\newblock
\urldef\tempurl%
\url{https://doi.org/10.1109/cvpr.2019.00293}
\showDOI{\tempurl}


\bibitem[Tan and Le(2019)]%
        {tan2019efficientnet}
\bibfield{author}{\bibinfo{person}{Mingxing Tan} {and} \bibinfo{person}{Quoc
  Le}.} \bibinfo{year}{2019}\natexlab{}.
\newblock \showarticletitle{Efficientnet: Rethinking model scaling for
  convolutional neural networks}. In \bibinfo{booktitle}{\emph{International
  conference on machine learning}}. PMLR, \bibinfo{pages}{6105--6114}.
\newblock


\bibitem[Wang(2001)]%
        {wang2001security}
\bibfield{author}{\bibinfo{person}{Chenxi Wang}.}
  \bibinfo{year}{2001}\natexlab{}.
\newblock \bibinfo{booktitle}{\emph{A security architecture for survivability
  mechanisms}}.
\newblock \bibinfo{publisher}{University of Virginia}.
\newblock


\bibitem[Wang(2021)]%
        {tflite2onnx}
\bibfield{author}{\bibinfo{person}{Zhenhua Wang}.}
  \bibinfo{year}{2021}\natexlab{}.
\newblock \bibinfo{booktitle}{\emph{{tflite2onnx - Convert TensorFlow Lite
  models to ONNX}}}.
\newblock
\urldef\tempurl%
\url{https://github.com/jackwish/tflite2onnx}
\showURL{%
\tempurl}


\bibitem[Wei et~al\mbox{.}(2020)]%
        {wei2020leaky}
\bibfield{author}{\bibinfo{person}{Junyi Wei}, \bibinfo{person}{Yicheng Zhang},
  \bibinfo{person}{Zhe Zhou}, \bibinfo{person}{Zhou Li}, {and}
  \bibinfo{person}{Mohammad~Abdullah Al~Faruque}.}
  \bibinfo{year}{2020}\natexlab{}.
\newblock \showarticletitle{Leaky dnn: Stealing deep-learning model secret with
  gpu context-switching side-channel}. In \bibinfo{booktitle}{\emph{2020 50th
  Annual IEEE/IFIP International Conference on Dependable Systems and Networks
  (DSN)}}. IEEE, \bibinfo{pages}{125--137}.
\newblock
\urldef\tempurl%
\url{https://doi.org/10.1109/dsn48063.2020.00031}
\showDOI{\tempurl}


\bibitem[Wroblewski(2002)]%
        {wroblewski2002general}
\bibfield{author}{\bibinfo{person}{Gregory Wroblewski}.}
  \bibinfo{year}{2002}\natexlab{}.
\newblock \showarticletitle{General method of program code obfuscation}.
\newblock  (\bibinfo{year}{2002}).
\newblock


\bibitem[Wu et~al\mbox{.}(2024)]%
        {wu2024concealing}
\bibfield{author}{\bibinfo{person}{Jing Wu}, \bibinfo{person}{Munawar Hayat},
  \bibinfo{person}{Mingyi Zhou}, {and} \bibinfo{person}{Mehrtash Harandi}.}
  \bibinfo{year}{2024}\natexlab{}.
\newblock \showarticletitle{Concealing Sensitive Samples against Gradient
  Leakage in Federated Learning}. In \bibinfo{booktitle}{\emph{Proceedings of
  the AAAI Conference on Artificial Intelligence}}, Vol.~\bibinfo{volume}{38}.
  \bibinfo{pages}{21717--21725}.
\newblock


\bibitem[Wu et~al\mbox{.}(2020)]%
        {wu2020decision}
\bibfield{author}{\bibinfo{person}{Jing Wu}, \bibinfo{person}{Mingyi Zhou},
  \bibinfo{person}{Shuaicheng Liu}, \bibinfo{person}{Yipeng Liu}, {and}
  \bibinfo{person}{Ce Zhu}.} \bibinfo{year}{2020}\natexlab{}.
\newblock \showarticletitle{Decision-based universal adversarial attack}.
\newblock \bibinfo{journal}{\emph{arXiv preprint arXiv:2009.07024}}
  (\bibinfo{year}{2020}).
\newblock


\bibitem[Zhang et~al\mbox{.}(2022)]%
        {zhang2022investigating}
\bibfield{author}{\bibinfo{person}{Chaoning Zhang}, \bibinfo{person}{Philipp
  Benz}, \bibinfo{person}{Adil Karjauv}, \bibinfo{person}{Jae~Won Cho},
  \bibinfo{person}{Kang Zhang}, {and} \bibinfo{person}{In~So Kweon}.}
  \bibinfo{year}{2022}\natexlab{}.
\newblock \showarticletitle{Investigating Top-k White-Box and Transferable
  Black-box Attack}. In \bibinfo{booktitle}{\emph{Proceedings of the IEEE/CVF
  Conference on Computer Vision and Pattern Recognition}}.
  \bibinfo{pages}{15085--15094}.
\newblock


\bibitem[Zhou et~al\mbox{.}(2024a)]%
        {zhou2024dynamo}
\bibfield{author}{\bibinfo{person}{Mingyi Zhou}, \bibinfo{person}{Xiang Gao},
  \bibinfo{person}{Xiao Chen}, \bibinfo{person}{Chunyang Chen},
  \bibinfo{person}{John Grundy}, {and} \bibinfo{person}{Li Li}.}
  \bibinfo{year}{2024}\natexlab{a}.
\newblock \bibinfo{booktitle}{\emph{DynaMO: Protecting Mobile DL Models through
  Coupling Obfuscated DL Operators (0.1)}}.
\newblock
\urldef\tempurl%
\url{https://doi.org/10.5281/zenodo.13762398}
\showDOI{\tempurl}


\bibitem[Zhou et~al\mbox{.}(2024b)]%
        {zhou2024model}
\bibfield{author}{\bibinfo{person}{Mingyi Zhou}, \bibinfo{person}{Xiang Gao},
  \bibinfo{person}{Pei Liu}, \bibinfo{person}{John Grundy},
  \bibinfo{person}{Chunyang Chen}, \bibinfo{person}{Xiao Chen}, {and}
  \bibinfo{person}{Li Li}.} \bibinfo{year}{2024}\natexlab{b}.
\newblock \showarticletitle{Model-less Is the Best Model: Generating Pure Code
  Implementations to Replace On-Device DL Models}. In
  \bibinfo{booktitle}{\emph{Proceedings of the 33rd ACM SIGSOFT International
  Symposium on Software Testing and Analysis}} (Vienna, Austria)
  \emph{(\bibinfo{series}{ISSTA 2024})}. \bibinfo{publisher}{Association for
  Computing Machinery}, \bibinfo{address}{New York, NY, USA},
  \bibinfo{pages}{174–185}.
\newblock
\showISBNx{9798400706127}
\urldef\tempurl%
\url{https://doi.org/10.1145/3650212.3652119}
\showDOI{\tempurl}


\bibitem[Zhou et~al\mbox{.}(2023)]%
        {zhou2023modelobfuscator}
\bibfield{author}{\bibinfo{person}{Mingyi Zhou}, \bibinfo{person}{Xiang Gao},
  \bibinfo{person}{Jing Wu}, \bibinfo{person}{John Grundy},
  \bibinfo{person}{Xiao Chen}, \bibinfo{person}{Chunyang Chen}, {and}
  \bibinfo{person}{Li Li}.} \bibinfo{year}{2023}\natexlab{}.
\newblock \showarticletitle{ModelObfuscator: Obfuscating Model Information to
  Protect Deployed ML-Based Systems}. In \bibinfo{booktitle}{\emph{Proceedings
  of the 32nd ACM SIGSOFT International Symposium on Software Testing and
  Analysis}} (Seattle, WA, USA) \emph{(\bibinfo{series}{ISSTA 2023})}.
  \bibinfo{publisher}{Association for Computing Machinery},
  \bibinfo{address}{New York, NY, USA}, \bibinfo{pages}{1005–1017}.
\newblock
\showISBNx{9798400702211}
\urldef\tempurl%
\url{https://doi.org/10.1145/3597926.3598113}
\showDOI{\tempurl}


\bibitem[Zhou et~al\mbox{.}(2024c)]%
        {zhou2024investigating}
\bibfield{author}{\bibinfo{person}{Mingyi Zhou}, \bibinfo{person}{Xiang Gao},
  \bibinfo{person}{Jing Wu}, \bibinfo{person}{Kui Liu},
  \bibinfo{person}{Hailong Sun}, {and} \bibinfo{person}{Li Li}.}
  \bibinfo{year}{2024}\natexlab{c}.
\newblock \showarticletitle{Investigating White-Box Attacks for On-Device
  Models}. In \bibinfo{booktitle}{\emph{Proceedings of the IEEE/ACM 46th
  International Conference on Software Engineering}}. \bibinfo{pages}{1--12}.
\newblock


\bibitem[Zhou et~al\mbox{.}(2020)]%
        {zhou2020dast}
\bibfield{author}{\bibinfo{person}{Mingyi Zhou}, \bibinfo{person}{Jing Wu},
  \bibinfo{person}{Yipeng Liu}, \bibinfo{person}{Shuaicheng Liu}, {and}
  \bibinfo{person}{Ce Zhu}.} \bibinfo{year}{2020}\natexlab{}.
\newblock \showarticletitle{Dast: Data-free substitute training for adversarial
  attacks}. In \bibinfo{booktitle}{\emph{Proceedings of the IEEE/CVF Conference
  on Computer Vision and Pattern Recognition}}. \bibinfo{pages}{234--243}.
\newblock


\end{thebibliography}










\end{document}